\documentclass[twocolumn]{aastex631}

\usepackage{xspace}
\usepackage{graphicx}
\usepackage{amssymb}
\usepackage{natbib}
\usepackage{dcolumn}

\newcolumntype{d}[1]{D{.}{.}{#1}}

\newcommand{\oiii}{[O{\scshape iii}]\xspace}

\newcommand{\nii}{[N{\scshape ii}]\xspace}

\newcommand{\dfour}{D$_{\rm n} (4000)$\xspace}

\begin{document}

\title{Selecting Post-Starburst Galaxies Based on Star Formation History}

\author[0009-0002-7155-6180]{Sara Starecheski}
\affiliation{Department of Astronomy, University of Maryland, College Park, MD 20742, USA}
\affiliation{University of Illinois Urbana-Champaign Department of Astronomy, University of Illinois, 1002 W. Green St., Urbana, IL 61801, USA}

\author[0000-0002-4235-7337]{K. Decker French}
\affiliation{University of Illinois Urbana-Champaign Department of Astronomy, University of Illinois, 1002 W. Green St., Urbana, IL 61801, USA}

\author[0000-0002-5877-379X]{Vicente Villanueva}
\affiliation{Instituto de Estudios Astrofísicos, Facultad de Ingeniería y Ciencias, Universidad Diego Portales, Av. Ejército Libertador 441, 8370191 Santiago, Chile} 
\affiliation{Millennium Nucleus for Galaxies, MINGAL}

\author[0000-0001-6444-9307]{Seb\'astion F. S\'anchez}\affiliation{Instituto de Astronom\'ia, Universidad Nacional Auton\'oma de M\'exico, A.P. 106, Ensenada 22800, BC, M\'exico}\label{uname}\affiliation{Instituto de Astrof\'\i sica de Canarias, La Laguna, Tenerife, E-38200, Spain \label{iac}}

\author[0000-0002-7759-0585]{Tony Wong}
\affiliation{University of Illinois Urbana-Champaign Department of Astronomy, University of Illinois, 1002 W. Green St., Urbana, IL 61801, USA}

\author[0000-0003-1535-4277]{Margaret E. Verrico}
\affiliation{University of Illinois Urbana-Champaign Department of Astronomy, University of Illinois, 1002 W. Green St., Urbana, IL 61801, USA}
\affiliation{Center for AstroPhysical Surveys, National Center for Supercomputing Applications, 1205 West Clark Street, Urbana, IL 61801, USA}

\author[0000-0002-8432-3362]{Alex Green}
\affiliation{University of Illinois Urbana-Champaign Department of Astronomy, University of Illinois, 1002 W. Green St., Urbana, IL 61801, USA}

\author[0000-0002-6582-4946]{Akshat Tripathi}
\affiliation{University of Illinois Urbana-Champaign Department of Astronomy, University of Illinois, 1002 W. Green St., Urbana, IL 61801, USA}

\author[0009-0004-5807-9142]{Keaton Donaghue}
\affiliation{Department of Astronomy, University of Maryland, College Park, MD 20742, USA}

\begin{abstract}
Post-StarBurst (PSB) galaxies are galaxies that have undergone a large burst of star formation followed by rapid quenching. Understanding their properties as a population can help us better understand how galaxies evolve to quiescence. This project aims to use Star Formation History (SFH) measurements from the Integral Field Spectroscopy (IFS) surveys MaNGA, CALIFA, and AMUSING++ processed with the Pipe3D analysis pipeline in order to select PSB galaxies as well as PSB regions in galaxies. Most PSB selection methods use cutoffs determined by spectral features, but in this work we introduce a new PSB selection method based directly on the property we are most interested in; inferred SFHs. IFS data allows us to probe a galaxy's star formation on a spatially resolved scale, enabling us to examine the size, shape, and location of PSB regions within a galaxy. We select 107 PSB galaxies, only 7 of which are among known PSBs selected by other methods. Unlike traditional PSB selection methods, our approach is not biased against Active Galactic Nuclei (AGN). Despite this, we still find no evidence for a significant Seyfert 2 PSB population, suggesting that strong AGN activity is uncommon throughout the PSB phase. Our spatially-resolved SFH selection identifies a wide range of galaxies, including globally quiescent elliptical galaxies with centrally-concentrated PSB spaxels, galaxies with ring-like PSB spaxels and a preference for inside-out age gradients (contrary to what has previously been observed in the literature), and galaxies with widespread PSB regions that have significant star formation elsewhere in the galaxy.
\end{abstract}

\section{Introduction} \label{sec:intro}

Galaxies can be broken up into two broad categories: blue disky galaxies with young stellar populations that are still forming stars at a significant rate, and red elliptical galaxies with old stellar populations that have entered a state of quiescence and have ceased significant star formation \citep{2004ApJ...600..681B, 2007ApJ...665..265F, 2001AJ....122.1861S, 2011MNRAS.415...32S, 2011ApJ...742...96W}. The transition between these two stages of galaxy evolution, or ``quenching," is not well understood. Quenching can look differently and occur in different spatial regions of a galaxy depending on the underlying mechanisms behind it. For example, if quenching occurs because of environmental effects such as starvation \citep{1980ApJ...237..692L}, or ram pressure stripping \citep{1972ApJ...176....1G}, we are more likely to see the impact of this on the outskirts of the galaxy. Stellar feedback such as supernova winds \citep{10.1111/j.1365-2966.2010.16872.x} is unlikely to result in total quiescence, but may quench specific regions of the galaxy. In the case of AGN feedback halting star formation \citep{1998A&A...331L...1S, 10.1111/j.1365-2966.2005.09238.x}, we are likely to see a nuclear deficit \citep{2021MNRAS.505L..46E}. In order to test for these effects, we must understand both when and where regions within galaxies quench. 

PSB galaxies represent an intermediate state between these two categories and are often found in the green valley \citep{2012MNRAS.420.1684W, 2017ApJ...843....9A}. PSB galaxies have recently experienced a large burst of star formation on a short timescale (a few tens to a few hundreds of Myr, e.g., \citealt{2018ApJ...862....2F}), but currently display low Star Formation Rates (SFRs), suggesting that they are undergoing quenching.
Studying PSB galaxies allows us to probe this transitional stage and gain a deeper understanding into the processes that drive galaxy evolution and how this stage of galaxy evolution depends on global physical properties \citep{2021PASP..133g2001F}. Thus, accurately identifying PSB galaxies on a large scale is essential for creating a large sample that can be studied as a population. 

There are many different ways to select PSB galaxies using their spectral features, which map to the star formation histories we aim to select for. For example, E+A (or K+A) galaxies are selected using a combination of spectral emission and absorption features \citep{1983ApJ...270....7D, 1987MNRAS.229..423C, 1996ApJ...466..104Z, 2003PASJ...55..771G, 2018ApJ...862....2F}. H$\alpha$ emission is associated with the presence of O stars: short-lived massive stars that are indicators of ongoing star formation. The Lick H$\delta_A$ index (measuring the strength of the H$\delta$ absorption line; \citealt{Worthey1997}) correlates to the presence of A stars, which have main sequence lifetimes $\sim1$ Gyr and are indicators of recent star formation. E+A galaxies are selected based on the presence of A stars, but not O stars (H$\delta_A$ $>$ 4 \AA\ in absorption, H$\alpha$ EW $<$ 3 \AA\ in emission) which indicates a recent burst that has not been replenished by ongoing star formation. Alternatively, PSB galaxies can be identified using a combination of features surrounding the Balmer break, for example the principal component analysis (PCA) method of \cite{2007MNRAS.381..543W, 2009MNRAS.395..144W} selects for PSBs using principal components corresponding roughly to the H$\delta_A$ and D4000 indices. Shocked POst-starburst Galaxies (SPOGs), another PSB galaxy sample, are alternatively identified using emission line ratios, selecting systems where lines are excited by shocks rather than ongoing star formation \citep{2016ApJS..224...38A}. These selection methods can be used with single-fiber data as in the studies referenced above, or for integral field spectroscopy surveys, with a focus on the radial distribution and spatial configurations of PSB spaxels within these galaxies \citep{2018MNRAS.480.2544R, 2019MNRAS.489.5709C, 2024ApJ...961..216C, 2025A&A...696A.206V}.

While these spectral index-based methods are widely used, there are a number of drawbacks in using them to select PSB galaxy samples. The E+A method described above selects a varying range of post-starburst ages, depending on the strength and duration of the recent burst. Older bursts are typically identified only for the cases where a large fraction of stellar mass is produced in the starburst, while weaker bursts are easier to identify when younger. \citet{2018ApJ...862....2F} showed that the distribution of PSB galaxies in the space of burst mass fraction and post-starburst age is a complex function of the typical burst SFRs, the typical durations of starbursts, and the visibility timescale of the different features (see their Figure 13). Furthermore, \citet{2018ApJ...862....2F} showed that neither H$\delta_A$ nor D4000 correlates well with the burst strength or age in all scenarios (see their Section 3.14). Ideally, PSB selection should be based directly on physical properties like post-burst age and burst mass fraction, but this has been limited in the past by a lack of reliable stellar population models applied to large spectroscopic samples. 

Here, we attempt to overcome these limitations by creating a set of selection criteria based directly on star formation histories. We do this using Integral Field Unit (IFU) spectroscopy, which allows us to observe spectra from different spatial regions of an extended object such as a galaxy.

We develop and test our selection criteria using three IFU surveys  of z $<$ 0.15 galaxies: MaNGA (Mapping Nearby Galaxies at Apache Point Observatory), CALIFA (Calar Alto Legacy Integral Field Area), and AMUSING++ (composed of galaxies from different MUSE (Multi Unit Spectroscopic Explorer) projects, with the majority from the All-weather MUSE Supernova Integral-field of Nearby Galaxies (AMUSING) survey). These surveys were chosen for their large sample sizes, and because all three have been processed with the Pipe3D analysis pipeline, which fits a detailed SFH for each spaxel. Obtaining the spatially resolved properties of galaxies for such a large sample size is extremely valuable because it connects the large scale of single aperture surveys with the detailed analysis we are able to achieve in studies of individual galaxies \citep{2012A&A...538A...8S}. This IFU technique is particularly useful in our case because this method allows us to map where and when sub-regions within galaxies have quenched, and to select galaxies experiencing quenching with both inside-out or outside-in radial trends, which describe whether star formation first shuts down in the center or in the outskirts, respectively.  

We describe the IFU data used in this work in \S\ref{sec:data}. We outline the methods used to identify PSBs based on their SFHs, and compare our sample to PSBs selected using other criteria in \S\ref{sec:methods}. In \S\ref{sec:results} we present the galaxies selected using our new methods, describe their global star formation properties, consider their morphologies and the spatial distributions of the PSB regions, consider their dominant sources of ionization, and characterize their age gradients. In  \S\ref{sec:discussion}, we search for differences between galaxies selected in each of our three IFU surveys, compare to other works, and discuss the implications of our findings for galaxy quenching. We conclude in \S\ref{sec:conclusion}. Our selection method can be used to identify additional PSB candidates in future surveys and the properties of our selected PSBs provide new insights into quenching and the mechanisms that drive galaxy evolution. 

\section{Data} \label{sec:data}
We use IFU data from the eCALIFA, MaNGA, and AMUSING++ surveys for our analysis. 

\subsection{The eCALIFA survey}
\label{eCALIFA}

We use the extended data release of the CALIFA survey \citep{2024RMxAA..60...41S}, comprised of 895 galaxies observed at the Calar Alto observatory, using the v500 setup only. These observations cover a redshift range of 0.0005 $<$ z $<$ 0.08, a wavelength range of 3745-7500 \text{\AA}, and a spectral resolution of R $\sim$ 850 \citep{2023MNRAS.526.5555S}. The CALIFA IFU is assembled in a hexagonal field of view, and uses a three pointing dithering scheme in order to account for gaps between the fibers and achieve full coverage.

\subsection{The MaNGA survey}
\label{MaNGA}

The MaNGA dataset was taken on the Sloan 2.5 meter telescope as part of the Sloan Digital Sky Survey (SDSS-IV). We use data release 17 \citep{2022ApJS..259...35A} consisting of 11,273 datacubes, of which 10,245 are unique galaxies. In total there are 10,220 galaxies with Pipe3D data \citep{article}. This survey covers a redshift range 0.01 $<$ z $<$ 0.15 \citep{2017AJ....154...28B}, and a wavelength range of 3600-10300 \text{\AA}, with a resolution of R $\sim$ 2000 \citep{2015ApJ...798....7B}. Like CALIFA, the MaNGA IFU is assembled in a hexagonal field of view, and has adopted a three pointing dithering scheme to account for gaps between the fibers \citep{2015AJ....149...77D}.

\subsection{The AMUSING++ sample}
\label{amusing++}

The AMUSING++ sample is observed with the MUSE integral-field spectrograph at the VLT \citep{2010SPIE.7735E..08B}. AMUSING++ is composed of galaxies from different MUSE projects, with the majority from the AMUSING survey. AMUSING++ comprises 635 galaxies with redshifts between 0.0002 $<$ z $<$ 0.1, a wavelength range of 4750 to 9300 \text{\AA}, and a resolution R between 1770 and 3590 \citep{2020AJ....159..167L}. Rather than using fibers, the AMUSING++ integral field spectrograph slices the image into 24 pieces and then sends each to its own IFU for processing the spectra.

\subsection{Using the Pipe3D Datacubes}
\label{Pipe3D}

All three of these surveys have been processed by pyPipe3D, an analysis pipeline that standardizes data products for large volumes of observational IFU data \citep{2022NewA...9701895L}. We obtain information about the stellar populations and emission lines of these galaxies from the pyPipe3D datacubes. Included in these datacubes are measurements of the fraction of the total luminosity formed at various ages throughout a galaxy's history \citep{2016RMxAA..52..171S}, and using stellar population models we calculate the fraction of the total mass for these same age bins. We can use these markers of star formation to identify recent bursts and low current SFRs in order to classify PSB galaxies based on their star formation history. Information obtained from the Balmer lines and other spectral indices, also included in Pipe3D, is used to diagnose the accuracy of our selection criteria and to further investigate the properties of our PSB sample. 

\section{Methods} \label{sec:methods}
\subsection{Summed Burst Criteria}
\label{Summed Burst Criteria}

We develop our SFH-based selection using two stages. First, we use the globally-averaged SFHs to define a selection with significant overlap with the traditional E+A selection, and to identify galaxies with global PSB SFHs (this section). In the next section (\S\ref{Percent Spaxel Burst Criteria}), we consider additional galaxies where a large fraction of spaxels meet the criteria we define here, typically due to widespread PSB regions in the outer galaxy.

\begin{figure*}
\includegraphics[width=\textwidth]{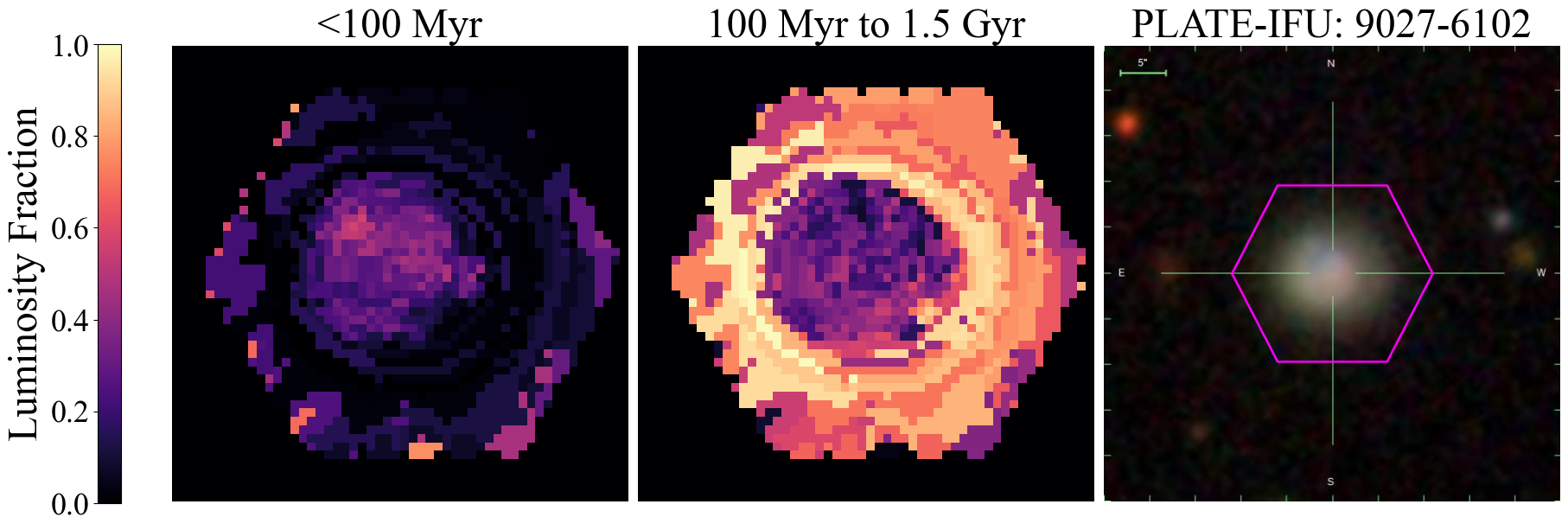}
\caption{An example of the spatially-resolved luminosity fractions from Pipe3D -- each spaxel is colored according to the fraction of the total luminosity that is from stars formed in the corresponding age bin. The left panel of the plot shows the most recent 100 Myrs, the middle panel shows between 100 Myr and 1.5 Gyr, and the right panel shows the SDSS $gri$ image of this galaxy with the MaNGA IFU field overlaid, from Marvin \citep{2019AJ....158...74C}. This galaxy is one of our Ring PSBs. Also note that our burst criteria is based off of mass fraction values rather than the luminosity fraction values, which are on average lower than the luminosity fraction values.
\label{fig:ifu}}
\end{figure*}

For our PSB selection, we aim to identify SFHs where the galaxy has had a burst of star formation within the last Gyr, that has since declined to a quiescent level. We define ``recent" star formation as occurring in the past 100 Myr, corresponding to the presence of O and B stars (typical lifetime of an intermediate B type star), and as the typical timescale probed by the blue/NUV region affecting the stellar population fits \citep{Kennicutt2012}. The 100 Myr - 1.5 Gyr range represents an intermediate age range corresponding to the presence of A stars. We use the pipe3D results to consider the resolved galaxy SFHs. The pyPipe3D datacubes contain spatially-resolved luminosity and mass fractions across 39 different age bins spanning 13.5 Gyr for MaNGA and CALIFA, and 14.1 Gyr for AMUSING++. These 39 raw Pipe3D age bins are not evenly spaced, but rather, are more tightly spaced at younger ages (visible in Figure \ref{fig:sfh}) such that for MaNGA and CALIFA, 14 of the raw Pipe3D age bins lie at ages $< 100$ Myr, and 12 fall between 100 Myr and 1.5 Gyr, while for AMUSING++ there are 17 and 13 bins in each age range, respectively. In Figure \ref{fig:ifu} we demonstrate the spatial trends in a galaxy for young stars (ages $< 100$ Myr) and intermediate age stars (ages 100 Myr - 1.5 Gyr). 

By using stellar libraries to model the stellar populations, we calculate the fraction of the total mass contributed by each stellar age range\footnote{The SSP we used for MaNGA is {\tt SDSS17Pipe3D\_v3\_1\_1.fits} \citep{2022ApJS..262...36S}, for eCALIFA is {\tt MaStar\_CB19.slog\_1\_5.fits.gz} \citep{2023MNRAS.526.5555S}, and for AMUSING++ is {\tt GSD01\_156.fits} \citep{2013A&A...557A..86C, 2022ApJS..262...36S}.}. To identify galaxies with low present-day SFRs we require the mass fraction formed in the past 100 Myr to fall below a given threshold, and to identify galaxies that have recently undergone a burst, we require the mass fraction to exceed a defined cut in the intermediate age bin. To mitigate potential noise in any individual age bin, we require this threshold be met at least two adjacent Pipe3D ages between 100 Myr and 1.5 Gyr. This selects for a large, short burst of recent star formation, in combination with a low present-day SFR.

To allow us to calibrate our SFH selection using previous selection criteria, we need to enable comparison between the spatially resolved pipe3D results and integrated, galaxy-wide spectroscopic tracers. We take the light-weighted average of the pipe3D results, over the entire spatial extent of each galaxy, to mimic the data we would obtain if it was taken with a single aperture instead of an IFU. This results in a single set of luminosity fractions and mass fractions as a function of age for each galaxy. We include only spaxels with signal to noise ratio (S/N) $>$ 10 in our light-weighted average, where the S/N is defined as the ratio of the median intensity flux to its standard deviation within the full analyzed wavelength range. Additionally, we include only galaxies with average S/N $>$ 10 in our sample. The total sample of galaxies left in each survey is included in Table \ref{tab:Table1}.

\begin{table}[t]
\centering
\caption{Galaxy Counts for Each Survey}
\begin{tabular}{l|r r r r}  
\hline\hline
& \multicolumn{1}{c}{MaNGA} & \multicolumn{1}{c}{CALIFA} & \multicolumn{1}{c}{MUSE} & \multicolumn{1}{c}{} \\ 
\hline
Original Sample & 10220 & 895 & 635\\
After S$/$N Cut & 9858 & 828 & 532 \\
Summed Burst PSB& 30 & 14 & 2\\
Percent Spaxel PSB & 31 & 24 & 6\\
\hline
\end{tabular}
\label{tab:Table1}
\end{table}

\begin{figure*}
\includegraphics[width=\textwidth]{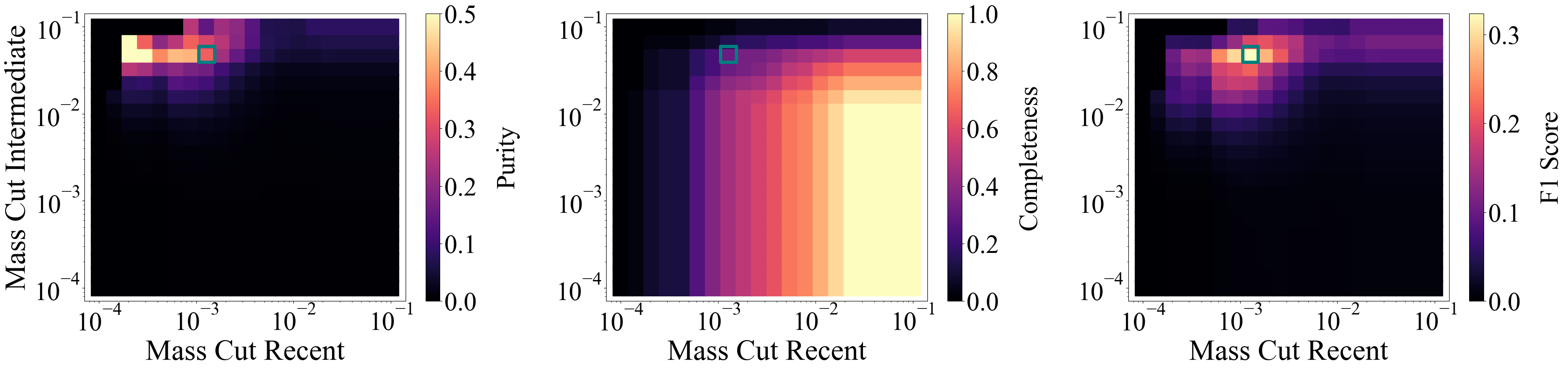}
\caption{Purity, completeness and F1 score for a range of different mass cuts in the recent age bin on the x axis and the intermediate age bin on the y axis. E+A selection criteria, H$\delta_A$ $>$ 4 \AA\ and H$\alpha$ $<$ 3 \AA\, were used as the benchmark by which we could measure the purity and completeness of our sample. We chose mass cuts that maximized the F1 score. These mass cuts -- $<0.13$\% in the recent age bin and $>4.8$\% in the intermediate age bin -- are highlighted on all three plots with a teal square. 
\label{fig:stats}}
\end{figure*}

We use traditionally-selected E+A galaxies to calibrate our SFH selection. For our E+A selection, we use the measurements from the pipe3D datacubes and require the light-weighted average value for H$\alpha$ EW to be $<$ 3 \AA\ in emission, and the light-weighted average value for Lick H$\delta_A$ to be $>$ 4 \AA\ in absorption. To calibrate our selection, we must balance the completeness (or recall) vs. purity (or precision) of our sample. A complete sample would include all galaxies meeting the E+A cut, while a pure sample would include only galaxies meeting this threshold. In Figure \ref{fig:stats}, we demonstrate the purity, completeness, and F1 score\footnote{A harmonic mean of the purity and completeness that provides a measure of accuracy.} achieved using a range of cuts for the recent and intermediate SFH mass cuts defined above. The selection with the highest F1 score achieves both a purity and completeness $>30$\%\footnote{Purity=33\%, Completeness=31\%, and F1=0.32.}. Thus, the selection method that we adopt here is:
\begin{itemize}
    \item Mass fraction formed in the most recent 100 Myr $<0.13$\%
    \item Mass fraction in at least two adjacent intermediate age bins between 100 Myr--1.5 Gyr $>4.8$\%
\end{itemize}

This selection criteria for PSBs will be known as our ``Summed Burst criteria". We performed this analysis on the MaNGA galaxies, and used these mass cuts to select PSBs in all three surveys. Since MaNGA contains the largest sample size, it provides us with the most accurate calibration, though we find consistent results if we consider CALIFA or AMUSING++. 

\begin{figure*}
\includegraphics[width=\textwidth]{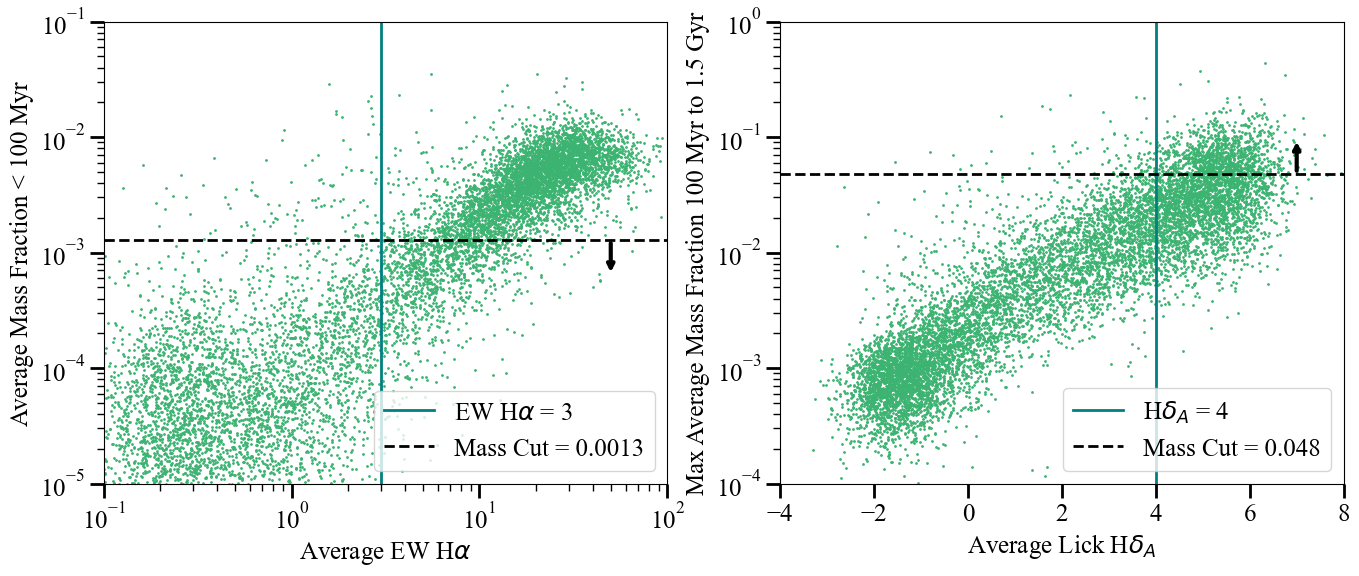}
\caption{(Left) EW H$\alpha$ vs the mass fraction in the corresponding recent age bin (the past 100 Myr) \citep{2014A&A...563A..49S, 2017MNRAS.469..151B}. The vertical line represents H$\alpha$ = 3, where H$\alpha$ $<$ 3 is one of the criteria for E+A PSBs, selecting for low sSFRs in the recent age bin. The horizontal line represents the mass cut in the recent age bin, where galaxies must have less than 0.13\% of their mass formed in this age bin to meet the Summed Burst criteria. (Right) Lick H$\delta_A$ vs the corresponding intermediate age bin (100 Myr to 1.5 Gyr). Note that a log scale is not used in this panel because the H$\delta_A$ index extends to negative values for quiescent galaxies due to a nearby unrelated feature in the spectra. The vertical line represents H$\delta_A$ = 4, where H$\delta_A$ $>$ 4 is one of the criteria for E+A PSBs, selecting for a burst of star formation in the intermediate age bin. The dashed horizontal line represents the mass cut in the intermediate age bin, where galaxies must have greater than 4.83\% of their mass formed in this age bin to meet the Summed Burst criteria. This data is shown for every galaxy in the MaNGA sample, and is taken from the pyPipe3D datacubes. Every point represents the light-weighted average of all spaxels in a single galaxy. There is a clear correlation between the star formation history and the emission lines we use to trace the presence of O and A stars. 
\label{fig:massfrac}}
\end{figure*}

\begin{figure*}
\includegraphics[width=\textwidth]{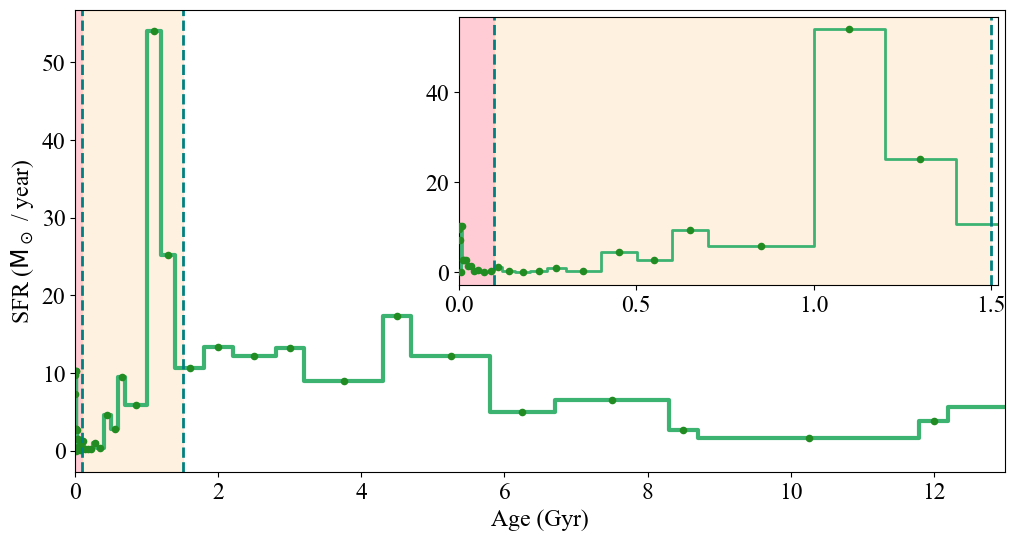}
\caption{Example of the SFH for one of our MaNGA Summed Burst PSBs (8616-12701) plotted over the past 13 Gyr in the main panel, and zoomed in over the past 1.5 Gyr in the inset panel. There is a large spike in star formation at approximately 1 Gyr, and low specific SFR at more recent ages. Teal vertical lines have been drawn at 100 Myr and 1.5 Gyr, and the corresponding regions have been shaded pink and peach respectively. The individual points represent values at the discrete ages.   
\label{fig:sfh}}
\end{figure*}

In Figure \ref{fig:massfrac}, we illustrate the extent to which our adopted SFH selection criteria correspond to the spectroscopic E+A selection cuts. 
Since EW H$\alpha$ should track star formation in the recent age bin \citep{Kennicutt1983, Kennicutt2012, 2014A&A...563A..49S, 2017MNRAS.469..151B}, and Lick H$\delta_A$ should track star formation in the intermediate age bin \citep{Worthey1997, 2018ApJ...862....2F}, we expect to see a correlation between H$\alpha$ and the mass fractions in the past 100 Myr. Likewise, we expect to see a correlation between H$\delta_A$ and the mass fractions from 100 Myr to 1.5 Gyr, since they should both be tracing the same intermediate stellar population. We plot these relationships for all MaNGA galaxies in our sample after the signal to noise cut. There is a clear correlation between the star formation history and the spectral lines we use to trace the presence of O and A stars, with a Pearson correlation coefficient of 0.4 for the lefthand plot, and a correlation coefficient of 0.9 for the righthand plot (p-value $\ll$ 0.05 for both). We find a fair amount of scatter, especially in H$\alpha$, as shown by its lower correlation coefficient, which indicates that our SFH selection may be sensitive to a different population of quenched galaxies. 

In total, 33 MaNGA galaxies, 14 CALIFA galaxies, and 3 AMUSING++ galaxies meet our Summed Burst criteria. To assess whether our selection has successfully identified galaxies with recent starbursts, we inspect the number of consecutive age bins that meet our intermediate mass fraction criterion. The majority of galaxies meeting our Summed Burst criteria only met the intermediate burst criterion at 2 adjacent ages (17 for MaNGA, 12 for CALIFA, 2 for AMUSING++). Only 3 galaxies met the criteria at 4 adjacent ages, and none met the criteria at 5 or more adjacent ages, indicating that we have been successful in selecting for shorter bursts. Figure \ref{fig:sfh} shows an example of the SFH for one of our Summed Burst PSBs. We can see that our SFH selection has successfully identified a SFH with a recent strong starburst and little to no recent star formation. The pipe3D SFHs for the sample of E+As in MaNGA typically show a peak SFR within the last 1.5 Gyr, and little to no recent star formation, consistent with what their positions would be if plotted on Figure \ref{fig:massfrac}. The E+As typically also have a secondary peak in star formation at older ages, when the older stellar population formed.

We plot the distributions of PSB spaxels (see Appendix \ref{app:1} for examples) for the galaxies we select above, highlighting spaxels that met this burst criteria. Based on visual inspection, we eliminate 3 galaxies from our MaNGA sample where there were doubts as to whether or not the majority of the PSB spaxels were the result of a foreground star contaminating our data. These have plate-IFUs of 11967-6102, 7958-6102, and 8615-12705. Note that in the Pipe3D datacubes, a mask is automatically applied to blank out foreground stars. For most galaxies these masks work quite well. However, our methods are especially susceptible to foreground star contamination because A star contamination can mimic PSB spaxels to our selection criteria. If we were to remove the PSB spaxels that we suspect are from a foreground star, none of the galaxies we removed would have $\ge$ 24\% PSB spaxels -- the average percent of PSB spaxels for galaxies meeting the Summed Burst criteria. We also eliminate a galaxy from the AMUSING++ sample that is unusually small in the field of view compared to the rest -- less than 5" and consisting of only a few spaxels -- making it difficult to examine its spatially resolved properties. This was 2dFGRSTGN327Z157. As described in Table \ref{tab:Table1}, this leaves 30 MaNGA galaxies, 14 CALIFA galaxies, and 2 AMUSING++ galaxies meeting our Summed Burst criteria. These galaxies are listed in Table \ref{tab:Table2}. We note that the PSB fractions in each survey may be influenced by differing selection effects and volumes probed, which we discuss further in \S\ref{betweensurveys}. 

We also investigate using shorter timescales to select against ongoing star formation, as Pipe3d uses 14 bins within the last 100 Myr. This yielded similar results: for example, using 33 Myr (corresponding to the expected age of ionizing O stars) as our recent age bin produced a Pearson correlation coefficient of 0.407 in the space of H$\alpha$ vs the average mass fraction formed in the recent age bin (compared to 0.404 in Figure \ref{fig:massfrac} when the recent age bin is set to be 100 Myr). Thus, the choice of recent age bin does not make a significant difference in the level of scatter we observe in H$\alpha$. Additionally, galaxies that quenched in the past 33 Myr are a subset of those that quenched in the past 100 Myr, so using 100 Myr as our recent age bin should not exclude any galaxies from our selection. 

\subsection{Percent Spaxel Burst Criteria}
\label{Percent Spaxel Burst Criteria}

Spatially-resolved data allow us to select additional PSB galaxies with widespread regions of spaxels that meet our PSB criteria, even if the spatially-integrated SFHs do not meet our Summed Burst criteria. In order to choose a threshold, we calculate the average percentage of PSB spaxels in each of the galaxies meeting the Summed Burst criteria, excluding spaxels that do not meet the S/N cut. The average percentage of PSB spaxels for galaxies meeting the Summed Burst criteria is 24\% for MaNGA, 18\% for CALIFA, and 51\% for AMUSING++. For galaxies not meeting the Summed Burst criteria, these values are 1\%, 5.4\%, and 4.4\% respectively. Note that these percentages were taken after the outliers were eliminated from the Summed Burst selection. We used the MaNGA results to calibrate the threshold for the other 3 surveys, adopting 24\% as our threshold. Any galaxies that do not meet our Summed Burst criteria, but do meet or exceed this threshold for PSB spaxels are considered part of our ``Percent Spaxel Burst criteria." In total, 31 MaNGA galaxies, 26 CALIFA galaxies, and 8 AMUSING++ galaxies meet these criteria. Two of the galaxies meeting these criteria for CALIFA were eliminated due to concerns about foreground star contamination. These were MGC+07-05-39 and UGC8688. Once again, two of the galaxies meeting the AMUSING++ criteria were eliminated due to being unusually small in our field of view: ASASSN-15lh and ESO379-24. The numbers of galaxies selected using this Percent Spaxel PSB selection method are included in Table \ref{tab:Table1}. The names of the galaxies meeting the Percent Spaxel Burst criteria are listed in Table \ref{tab:Table2}.

\begin{table*}
\centering
\caption{PSB Galaxies Selected by Each Criteria}
\begin{tabular}{c c c c c c c}  
\hline\hline
& \multicolumn{2}{c}{MaNGA} & \multicolumn{2}{c}{CALIFA} & \multicolumn{2}{c}{AMUSING++} \\
& Summed & \% Spax. & Summed & \% Spax. & Summed & \% Spax. \\ 
\hline
1 & 10840-9102 & 10222-9102 & 2MASXJ22532475 & AGC210538 & NGC7253 & NGC0924 \\
2 & 11009-3703 & 10505-6104 & CGCG008-023 & ARP180 & NGC7582 & NGC1068 \\
3 & 11940-6101 & 10510-12703 & IC3586 & ASASSN15cb &  & NGC4424 \\
4 & 11941-6101 & 10843-12703 & NGC0515 & CGCG97063a &  & ESO358-59 \\
5 & 11942-6101 & 11748-9102 & NGC6261 & CGCG97121 &  & ESO467-62 \\
6 & 11952-3703 & 11863-12702 & PGC0213858 & IC1014 &  & NGC2915 \\
7 & 11976-1901 & 11951-12703 & PGC213672 & IC1078 &  &  \\
8 & 12067-3701 & 11956-9102 & PGC37426 & IC2402 &  &  \\
9 & 12094-1902 & 11967-12701 & SN2006kf & IC3328 &  &  \\
10 & 12652-12701 & 11975-12703 & SN2012ij & NGC4211NED02\_0 &  &  \\
11 & 12673-12703 & 12082-9102 & SN2013N & NGC4390 &  &  \\
12 & 7495-6103 & 12088-6102 & UGC08234 & NGC4644 &  &  \\
13 & 8079-6101 & 12090-12703 & UGC08250 & NGC5682 &  &  \\
14 & 8570-12703 & 12483-3703 & VIIZw466b\_3 & PTF14aaf\_0 &  &  \\
15 & 8604-6101 & 12511-9102 &  & SN2001br &  &  \\
16 & 8616-12701 & 8139-12701 &  & SN2005bl &  &  \\
17 & 8625-6102 & 8155-12704 &  & SN2016hnk &  &  \\
18 & 8710-6104 & 8248-6104 &  & UGC02405 &  &  \\
19 & 8932-12705 & 8338-12704 &  & UGC08733 &  &  \\
20 & 8934-9101 & 8442-3704 &  & UGC10650 &  &  \\
21 & 8935-12701 & 8449-6102 &  & UGC112866 &  &  \\
22 & 8950-6101 & 8483-12702 &  & UGC1859 &  &  \\
23 & 8951-12703 & 8550-6102 &  & UGC5392 &  &  \\
24 & 8979-1902 & 8564-6103 &  & UGC8932 &  &  \\
25 & 9494-3701 & 8710-9102 &  &  &  &  \\
26 & 9494-6102 & 8932-9102 &  &  &  &  \\
27 & 9495-3702 & 9002-12705 &  &  &  &  \\
28 & 9872-12703 & 9027-6102 &  &  &  &  \\
29 & 9875-12703 & 9506-12701 &  &  &  &  \\
30 & 9877-3701 & 9889-6104 &  &  &  &  \\
31 &  & 9891-6102 &   &  &  &  \\
\hline
\end{tabular}
\label{tab:Table2}
\end{table*}

\subsection{Crossmatching With E+A, PCA, and SPOG PSBs}
\label{Crossmatching}

Our new PSB selection method identifies new PSB galaxies, as well as recovering some previously selected cases. \cite{2023arXiv230410419F} describe a sample of 5040 PSBs in SDSS, compiled from three overlapping sources: the E+A sample, the SPOG sample, and the PCA sample. These selection methods all use the SDSS single-fiber data, which is sensitive to the inner 1-2 kpc of the galaxy, in contrast to the IFU data we consider here. We cross-match our Summed Burst and Percent Spaxel Burst PSBs with these 5040 SDSS PSBs to determine if our selection represents a unique population of PSBs. Of these SDSS PSBs, 93 galaxies overlap with MaNGA\footnote{Note that the sample of known PSBs for MaNGA is composed of 93 galaxies, however, one PCA galaxy, PLATE-IFU 8454-6102, was eliminated by our requirement for the average signal to noise ratio of the galaxy to be greater than 10.}, 4 overlap with CALIFA, and 1 overlaps with AMUSING++ \footnote{These are the PCA galaxies ARP220, IC4526, SN2009cz, and UGC12633 from CALIFA, and ESO113-18 from AMUSING++, which is classified as both a PCA and an E+A.}. This makes sense as MaNGA is not only the largest sample of our three surveys, but CALIFA and AMUSING++ also cover different areas of the sky than SDSS. 

Of the 98 galaxies that overlap with our full sample, only 7 of them are selected by our SFH PSB criteria, for a total of 6 galaxies overlapping with our Summed Burst criteria and 1 overlapping with our Percent Spaxel Burst criteria. All seven of these are from MaNGA. From our Percent Spaxel Burst criteria, we find 1 PCA galaxy, and from our Summed Burst criteria we find 1 E+A galaxy and 5 that are classified as both PCA and E+A. These numbers reflect 0\% of the 19 SPOGs in the 98 galaxies, 8\% of the 83 PCA galaxies, and 24\% of the 26 E+A galaxies. 

The difference in galaxies we select compared to the SDSS single-fiber-selected samples is due to two effects. The first is the SFH-based selection method we develop earlier in this section, and the second is the use of IFU data as opposed to the single-fiber SDSS spectrum. The Summed Burst criteria method is the most similar to the single-fiber data methods, as the light-weighted average used in this method prioritizes the central portion of the galaxy. Thus, it is expected that this method will select a larger number of known PSBs compared to the Percent Spaxel method. The fact that we see the most overlap with the E+A sample is also expected, because the mass cuts in the recent and intermediate age bins were determined by maximizing the purity and completeness using the E+A criteria as a benchmark. This $\sim$7\% overlap confirms we are indeed selecting galaxies with PSB properties. However, the fact that the majority of our sample does not overlap with the traditionally selected PSBs tells us that we are selecting a unique population of PSB galaxies, with different properties than PSBs selected with previous methods.

\subsection{Visual Classification}
\label{Visual Classification}

Our SFH selection method allows us to identify regions within galaxies that are PSB and to identify whether the PSB spaxels are centrally concentrated or present in the outskirts of galaxies. Looking at the spatially resolved maps of PSB spaxels, we visually classified the distribution of the PSB spaxels in a similar style to the work done in \cite{2019MNRAS.489.5709C}. To aid in our visual classifications, we also pulled images of the galaxies from the Legacy Surveys \citep{2019AJ....157..168D}, and from DSS for areas of the sky that the Legacy Surveys did not cover. 

We classify the PSB spaxel morphologies as follows:

\begin{enumerate}
\item Ring: Galaxies classified as ``Ring" either displayed a clear, contiguous ring of PSB spaxels, or tended to have most PSB spaxels distributed around the edges of the field of view, with a clear hole in PSB spaxels in the middle.
\item Central: Galaxies classified as ``Central" displayed a clear clustering of PSB spaxels in the center of the galaxy. Note that some of this may be due to the galaxy being relatively small in the field of view.
\item Irregular: Galaxies classified as ``Irregular" did not display either of these characteristics, and tended to have a random distribution of PSB spaxels. Note that we tended to err on the side of classifying galaxies as Irregular if there was a reasonable level of uncertainty as to our classification.
\item Irregular + Companion: Galaxies classified as ``Irregular + Companion", a classification not included in \cite{2019MNRAS.489.5709C}, showed a second galaxy in the field of view, or obvious signs of a merger. In some cases, both galaxies contain PSB spaxels.
\end{enumerate}

An example of each of these distributions of PSB spaxels is shown in Figure \ref{fig:morph}. Maps of the distribution of PSB spaxels for all PSB galaxies are shown in Appendix \ref{app:1}, for completeness. The total number of galaxies falling into each of these classifications is summarized in Table \ref{tab:Table3}. 

\begin{figure*}
\includegraphics[width=\textwidth]{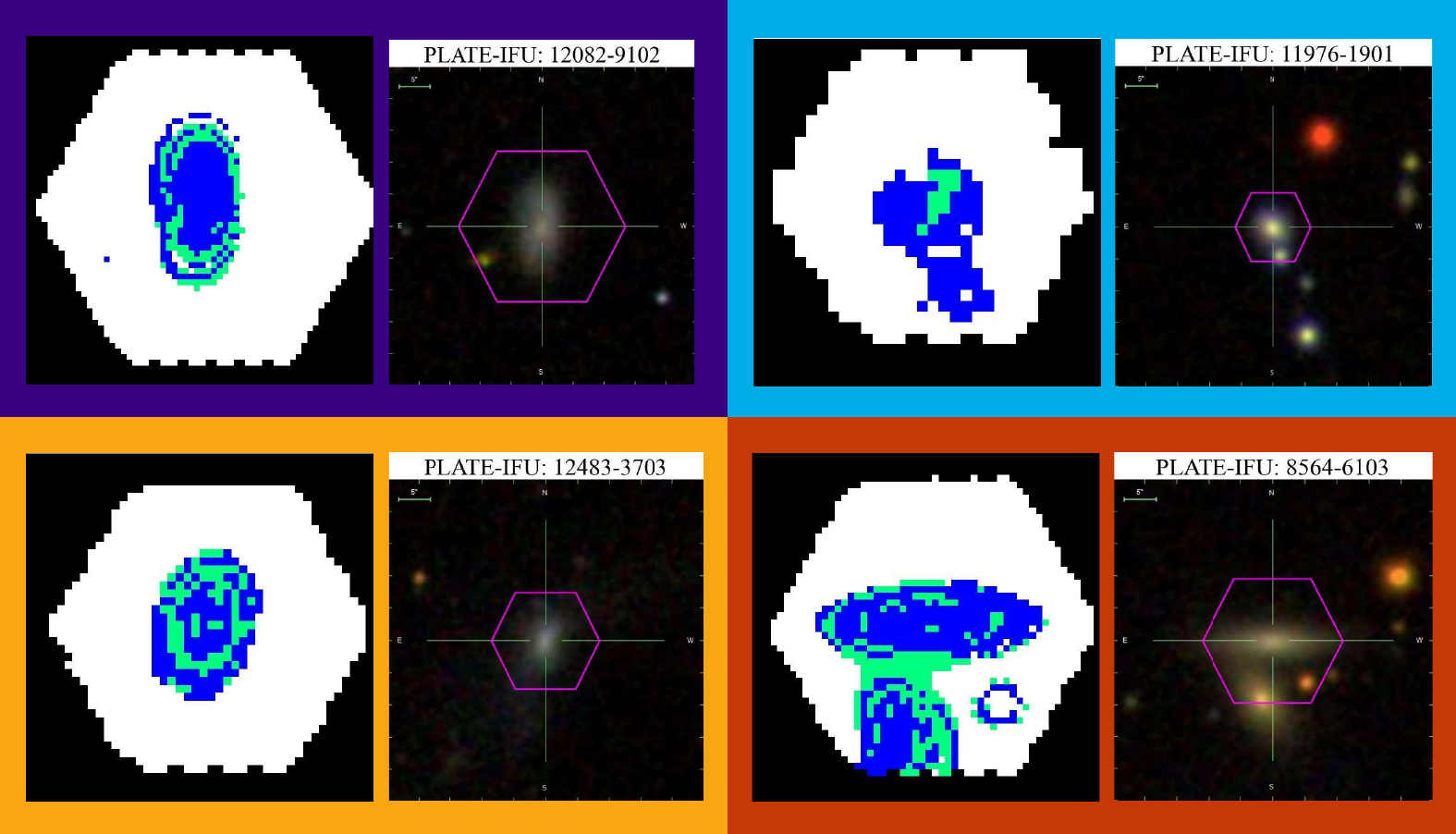}
\caption{Examples of a Ring (indigo), Central (blue), Irregular (orange), and Irregular + Companion (red) distribution of PSB spaxels from MaNGA. According to our selection criteria, PSB spaxels and non-PSB spaxels are shown in green and blue respectively. The limits of the hexagonal IFU are visible. In each panel, the leftmost column shows the IFU data after the signal to noise cut, and the rightmost column shows the image of the galaxy from Marvin \citep{2019AJ....158...74C}. The Ring galaxy shows all of its PSB spaxels in a continuous ring, with a clear hole in the center. The Central galaxy shows a concentration of PSB spaxels in the center. The Irregular galaxy does not fit into either of the above categories, and shows a random distribution of PSB spaxels throughout both the center and edges. The Irregular + Companion galaxy has some of its PSB spaxels coming from a nearby galaxy.}
\label{fig:morph}
\end{figure*}

\subsection{Galaxy Zoo Morphologies}
\label{Galaxy Zoo Morphologies}

In addition to the morphologies of the PSB spaxels, we also consider the morphology of the galaxy continuum light, as the presence of disk vs. elliptical morphologies, bars, tidal tails, and other features are strongly linked to the evolution of galaxies. We use Galaxy Zoo data release 2 \citep{2013MNRAS.435.2835W} classifications to consider the morphologies of the PSB galaxies in our selection. Galaxy Zoo\footnote{https://data.galaxyzoo.org/} is a citizen science project where participants classify not only whether galaxies are spirals or ellipticals, but also determine morphological features such as bars or mergers. We used the {\tt gz2\_class} data, which represents the morphology consensus. For cases where the PSBs have no match in Galaxy Zoo, we visually classify these galaxies using Legacy Survey \citep{2019AJ....157..168D} and DSS imaging. We use the Galaxy Zoo classification chart as a guide to ensure consistency.      

\section{Results} \label{sec:results}

In \S\ref{Summed Burst Criteria} and \ref{Percent Spaxel Burst Criteria} we have presented not only new selection criteria that can be used in future IFU surveys, but also a new sample of PSBs that is distinct from previous selections. Combining the PSBs selected from MaNGA, CALIFA, and AMUSING++, we select 46 PSBs with spatially-integrated SFHs that meet our selection criteria and 61 PSBs for which a sufficient number of spaxels meet our selection criteria (see Tables \ref{tab:Table1} and \ref{tab:Table2}), for a total of 107 PSBs. We now consider the global properties of these newly-selected PSBs to determine their likely evolutionary paths.

\subsection{Global Star Formation Properties}
\label{subsec:global} 

In order to compare the global star forming properties of these newly-selected PSBs to the broader galaxy population, we consider their locations on several common galaxy scaling relations. In each of these diagnostic plots, we use emission line data and spectral indices from the pyPipe3D datacubes, spatially averaged over each galaxy using light-weighted averaging, in the same manner as done for the SFH analysis above. We also investigate whether or not any trends in the global star-forming properties correlate with the four PSB spaxel morphologies introduced earlier. 

\begin{figure*}
\plottwo{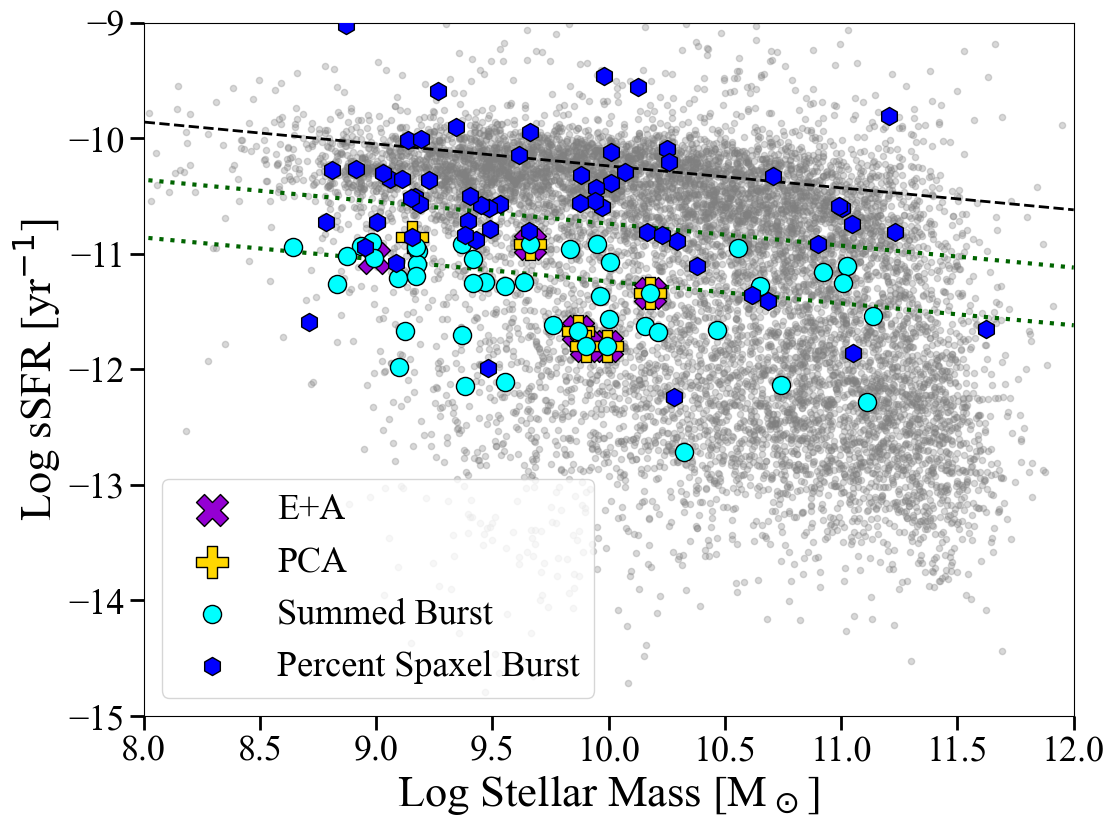}{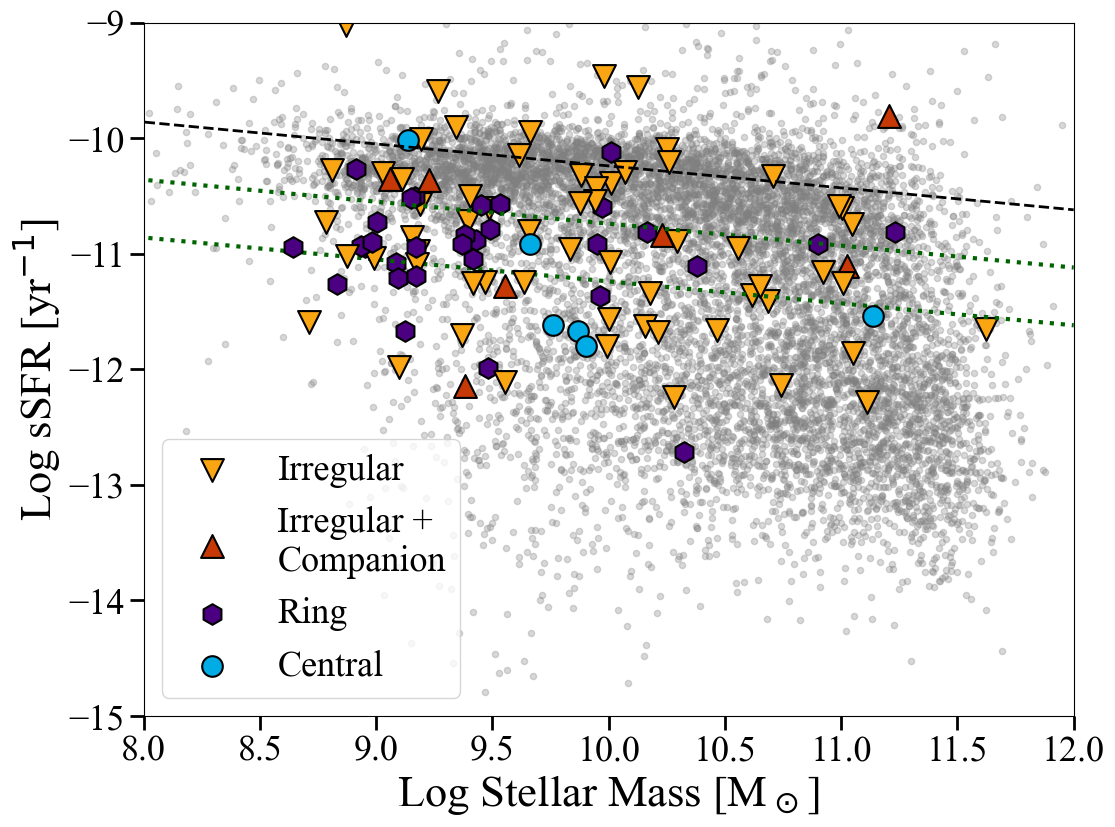}
\caption{Stellar mass vs sSFR for all galaxies in the MaNGA, CALIFA, and AMUSING++ surveys after the signal to noise cut. The star forming main sequence from  \cite{2016ApJ...821L..26C} is modeled by the dashed black line, and the dotted green lines mark the bounds for the green valley. In the left panel, we compare our Summed Burst PSBs to our Percent Spaxel Burst PSBs. In the right panel, we compare our four PSB spaxel distributions. Many of our Summed Burst PSBs and Centrals lie in the green valley or quenched regions of the plot, telling us that they have low rates of ongoing star formation. This is expected given that PSB galaxies are in a transitional state between star-forming and quiescence. Meanwhile, our Percent Spaxel Burst criteria PSBs skew higher in terms of specific SFR, more often lying in the star-forming region of the plot. This could be because the Percent Spaxel galaxies often have ongoing central star formation.
\label{fig:stellarmasssfr}}
\end{figure*}

We first consider the global star formation properties by plotting the specific SFR (sSFR) vs the stellar mass in Figure \ref{fig:stellarmasssfr}. We calculate the specific SFR using the mass fraction formed in the most recent 100 Myr from the Pipe3D datacubes. 
We obtain stellar masses for galaxies in each survey using the SDSS galspec analysis for MaNGA \citep{2003MNRAS.341...33K, 2004MNRAS.351.1151B, 2004ApJ...613..898T}, the \citet{2024RMxAA..60...41S} catalog for CALIFA ({\tt eCALIFA.pyPipe3D.fits}), and the \citet{2020AJ....159..167L} catalog for AMUSING++ ({\tt ajab7848t3\_mrt.txt}). We model our region for the green valley after \cite{2024ApJ...962...88V}, who use a definition of the SFMS (star forming main sequence) from \cite{2016ApJ...821L..26C}, and adopt a green valley region from -1 to -0.5 dex below the main sequence. This selection is typical of other green valley cuts \citep{2020A&A...644A..97C, 10.1093/mnras/stu327, Chang_2015, annurev:/content/journals/10.1146/annurev-astro-081915-023441, 2018RMxAA..54..217S}. 

Both the Summed Burst criteria and the Percent Spaxel Burst criteria have some scatter, but we observe a clear distinction between our two sets of selection criteria. We find that many (19/46) of our Summed Burst PSBs lie in the green valley, with the rest falling below in the quiescent region. This is what we expect, given that PSB galaxies are in a transitional state between star-forming and quiescence. While the Summed Burst PSBs that do not lie in the green valley primarily scatter into the quiescent region of the plot, PSBs meeting our Percent Spaxel Burst criteria skew higher in terms of specific SFR, more often lying in the star-forming region of the relation. The Percent Spaxel Galaxies must have star formation elsewhere in the galaxy, outside of the PSB spaxels, and significant enough that the average galaxy spectrum does not meet our Summed Burst criteria. A KS test between these two distributions in log sSFR yields a p-value $\ll$ 0.05, indicating that they are drawn from different distributions. Meanwhile, a KS test in log stellar mass yields a p-value of 0.995, indicating that these two samples are indistinguishable in stellar mass. 

Most of our Centrals (5/6) are in the green valley or quenched regions of the plot, below the star forming main sequence, indicating that they have low rates of ongoing star formation. A plurality of the Ring galaxies are found in the green valley (15/31), and in the half of the plot corresponding to lower stellar masses (25/31). The trend in stellar mass may be because galaxies with lower stellar masses are often closer due to observational biases, which could make it easier to resolve the ring distribution. Another possibility is that the lower stellar masses seen in our Ring PSB galaxies could be caused by fundamentally different quenching patterns. We explore this possibility further in Section \ref{subsec:radial}. The appreciable scatter of all types of PSB spaxel distributions shows that we are selecting PSB galaxies in various stages of quenching. Once again, the presence of galaxies with PSB regions in the star forming main sequence indicates that despite the quenched regions, star formation can continue elsewhere in the galaxy. For the total galaxy sSFR to be on the main sequence, despite the quenched regions we identify as PSB, the star-forming regions may be in the brighter or more highly weighted regions of the galaxy. 

\begin{figure*}
\plottwo{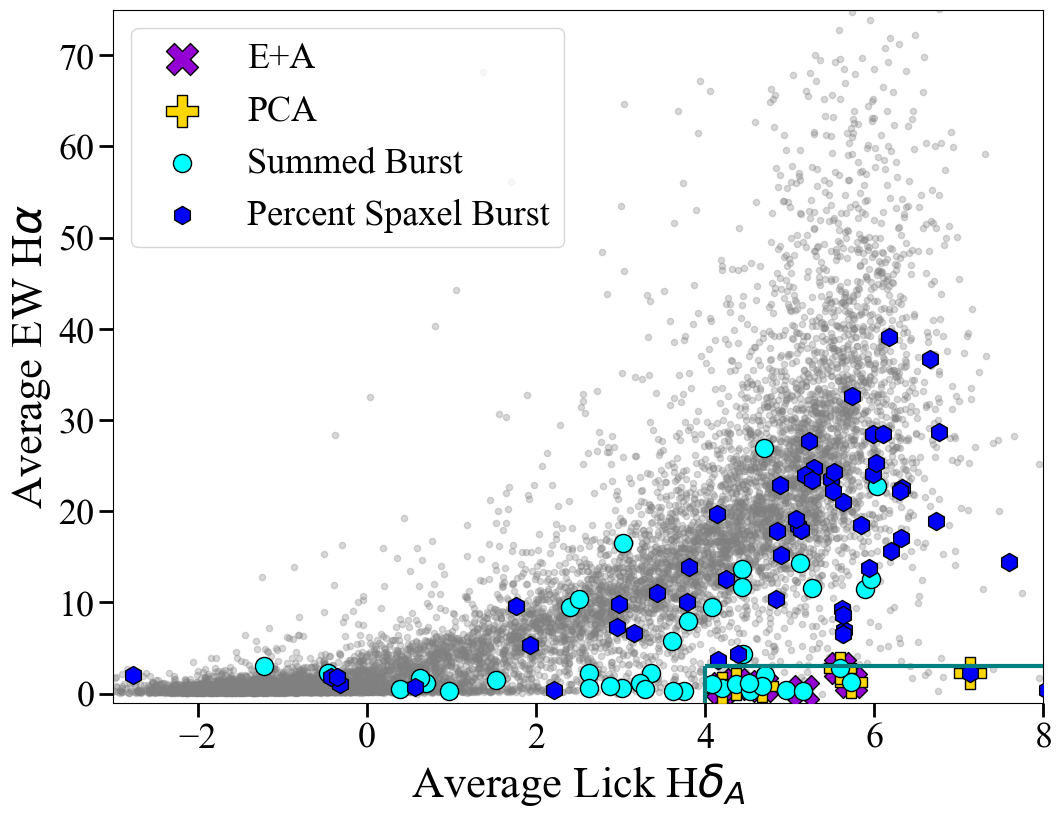}{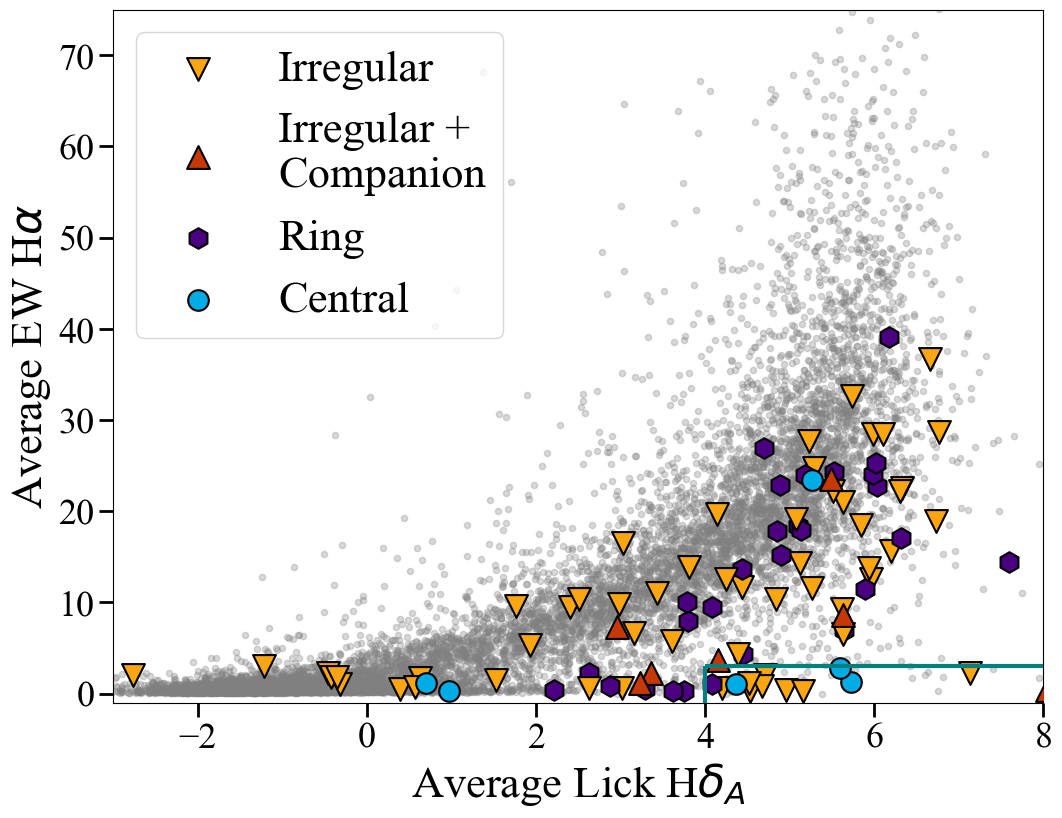}
\caption{Average Lick H$\delta_A$ vs. average EW H$\alpha$ for all galaxies in the MaNGA and CALIFA surveys after the signal to noise cut. Note that a log scale is not used in this plot because we have some negative values. A thin teal box has been drawn to designate the area of the plot where H$\alpha$ $<$ 3 and H$\delta_A$ $>$ 4. Galaxies that meet one of our two sets of criteria and are also in the sample of the 5040 SDSS PSBs described in \cite{2023arXiv230410419F} have a purple $\times$ appearing under them if they are in the E+A sample, a yellow + if they are in the PCA sample, and both an $\times$ and a + if they are in more than one of these samples. Note that we do not have any SPOGs that overlap with our sample. Many of our Summed Burst PSBs and Centrals lie in the H$\alpha$ $<$ 3 region of the plot, indicating that they are often quiescent. Conversely, many of our Percent Spaxel Burst PSBs, Rings, and Irregulars have ongoing star formation in other regions of the galaxy.
\label{fig:hahd}}
\end{figure*}

E+A galaxies are selected using Lick H$\delta_{\rm A}$ and EW H$\alpha$, thus, in order to compare our SFH selected PSBs to previous selections, we plot all MaNGA and CALIFA galaxies in this space in Figure \ref{fig:hahd}. This plot should be thought of as an alternate diagnostic of star formation, where H$\alpha$ is a measure of whether or not there is ongoing star formation, and H$\delta_A$ traces star formation on intermediate timescales. In the space of H$\delta_A$ absorption vs. H$\alpha$ emission, E+A PSBs occupy the bottom right corner where H$\delta_A$ $>$ 4 \AA\ and H$\alpha$ $<$ 3 \AA, (teal box in Figure \ref{fig:hahd}). This region contains galaxies with a strong recent burst and little ongoing star formation. All overlap with the SDSS PSBs is in this corner, as discussed Section \ref{Crossmatching}. Six of the 7 SDSS PSBs that overlap with our sample are classified as E+As. However, overlap with the SDSS PSBs is still a minority of our sample, and many of our selected PSBs across both sets of criteria have values for H$\alpha$ and H$\delta_A$ more consistent with the bulk of the non-PSB galaxies. This is the case because there is scatter between H$\delta_A$ and the mass fraction in the 100 Myr - 1.5 Gyr age range, and scatter between H$\alpha$ EW and the mass fraction in the most recent 100 Myr, visible in Figure \ref{fig:massfrac}. 

It is important to note that in AMUSING++, Lick H$\delta_A$ is outside of the wavelength range of the MUSE instrument. For this reason, AMUSING++ is not included in any diagnostic plots that use H$\delta_A$. However, this presents us with an important use-case for our selection criteria, as other selection methods that require the H$\delta_A$ index will miss the PSB galaxies in surveys like AMUSING++ that do not cover this wavelength range. 

Interestingly, many of our Summed Burst PSBs lie in the H$\alpha$ $<$ 3 \AA\ (quiescent) region of the plot, with scatter in H$\delta_A$. It is possible that this is due to older and weaker bursts. Conversely, many of our Percent Spaxel Burst PSBs lie in the expected space for H$\delta_A$, with high H$\delta_A$ values usually greater than 4 \AA. However, the scatter in H$\alpha$ is more pronounced, with very few lying in the H$\alpha$ $<$ 3 \AA\ region of the plot, placing our Percent Spaxel Burst PSBs with the bulk of star-forming galaxies. This difference could be because the Percent Spaxel Burst PSBs often have ongoing star formation in other regions of the galaxy, whereas the Summed Burst PSBs are more often quiescent. 

In total, 3 Centrals, 8 Irregulars, 1 Ring, and 1 Irregular + Companion meet the traditional E+A criteria. Notably, many of our Centrals and Irregular + Companions have low H$\alpha$ values, whereas our Ring galaxies typically have high H$\alpha$ values. For our Irregular + Companion galaxies, the low H$\alpha$ EWs could be due to dust obscuration rather than low amounts of ongoing star formation. Indeed, we often see regions of high extinction as traced by the Balmer decrement in our Irregular + Companion galaxies. For many of the Ring and Irregular cases, there is significant ongoing star formation elsewhere in the galaxy outside of these PSB regions. Three of the six Centrals overlap with the SDSS PSBs selected using traditional methods. This is likely due to the fact that the bright center dominates the light in the single-fiber selection, as well as the light-weighted average over the IFU. The PSB galaxies that overlap with the SDSS PSBs are all Central or Irregular galaxies -- none of the Ring PSBs are selected by the single-fiber selection methods.   

\begin{table}
\centering
\caption{Number of PSBs in each PSB spaxel morphology category, by selection method and parent survey}
\begin{tabular}{l|c c c c}  
\hline\hline
 & MaNGA & CALIFA & AMUSING++ \\ 
\hline
\hline
\multicolumn{4}{c}{{Summed Burst Criteria}} \\ 
\hline
Ring & 12 & 1 & 0\\
Central & 3 & 2 & 0\\
Irregular & 13 & 11 & 1\\
Irregular + Companion & 2 & 0 & 1\\
\hline
\multicolumn{4}{c}{{Percent Spaxel Burst Criteria}} \\ 
\hline
Ring & 15 & 0 & 3 \\
Central & 0 & 1 & 0\\
Irregular & 13 & 21 & 3\\
Irregular + Companion & 3 & 2 & 0\\
\hline
\end{tabular}
\label{tab:Table3}
\end{table}

In Table \ref{tab:Table3}, we show the number of PSBs for each PSB spaxel morphology, broken down by PSB selection method and parent survey. We see that the majority of Centrals come from the Summed Burst criteria. This is what we expect, seeing as our Summed Burst galaxies should be centrally-quenched on average if their light-weighted average SFHs meet our PSB criteria. For our Percent Spaxel Burst criteria on the other hand, there is a possibility of ongoing star formation in the centers, given that their light-weighted average star formation histories do not meet our PSB criteria. This allows the starburst and the subsequent quenching to happen elsewhere in the galaxy, so long as it happens over an appreciable enough fraction of the galaxy. We find examples of star formation occurring both in the central regions and in the outskirts of our PSBs. Many of our Percent Spaxel PSBs display widespread star-forming regions present throughout the galaxy, whereas many of our Summed Burst PSBs exhibit little to no star formation at all.   

We consider whether or not the differences we observe between the Summed Burst PSBs and the Percent Spaxel PSBs could be caused by differences between the Ring, Central, Irregular, and Irregular + Companion PSB spaxel distributions. We expect that the central region of our PSB galaxies should dominate the light-weighted averages we take for our Summed Burst criteria, and will thus produce similar properties between our Central and Summed Burst PSBs. In Figure \ref{fig:stellarmasssfr}, we see that both the Centrals and the Summed Burst galaxies tend to be quiescent. A KS test performed between the Centrals and the Summed Burst PSBs yields a p-value of 0.545 for log stellar mass and a p-value of 0.636 for log sSFR. This indicates that we cannot disprove the idea that these samples are from the same parent population. Turning to Figure \ref{fig:hahd}, we see a similar correlation: both the Summed Burst PSBs and the Centrals are often quiescent compared to the Ring and Irregular galaxies, which frequently scatter to higher H$\alpha$ values. Once again, a KS test performed between the Centrals and the Summed Burst PSBs confirms that we cannot distinguish between these distributions, with a p-value of 0.530 for H$\delta_A$ and a p-value of 0.756 for H$\alpha$. Although the relatively small number of Centrals we observe makes it difficult to draw many conclusions, our Centrals often follow trends that mimic those of our Summed Burst PSBs.

\subsection{Morphologies}
\label{subsec:galzoo}

\begin{figure*}
\plottwo{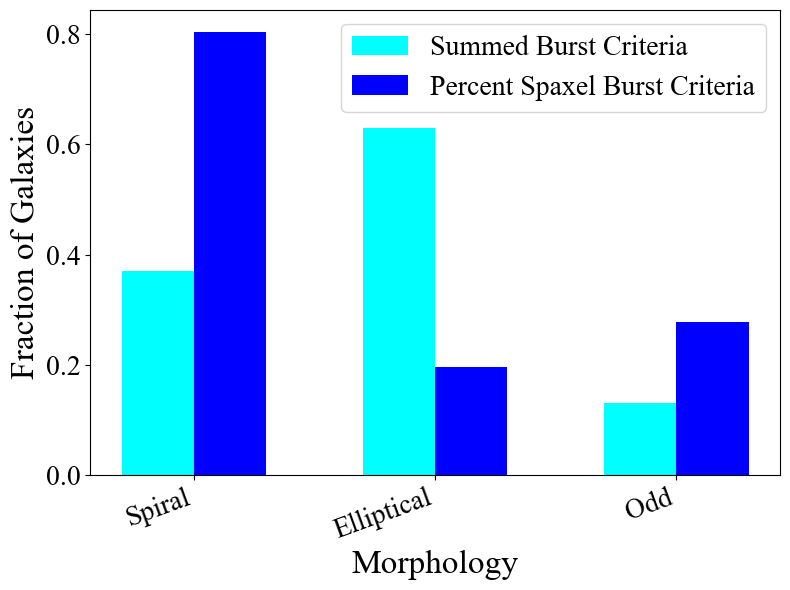}{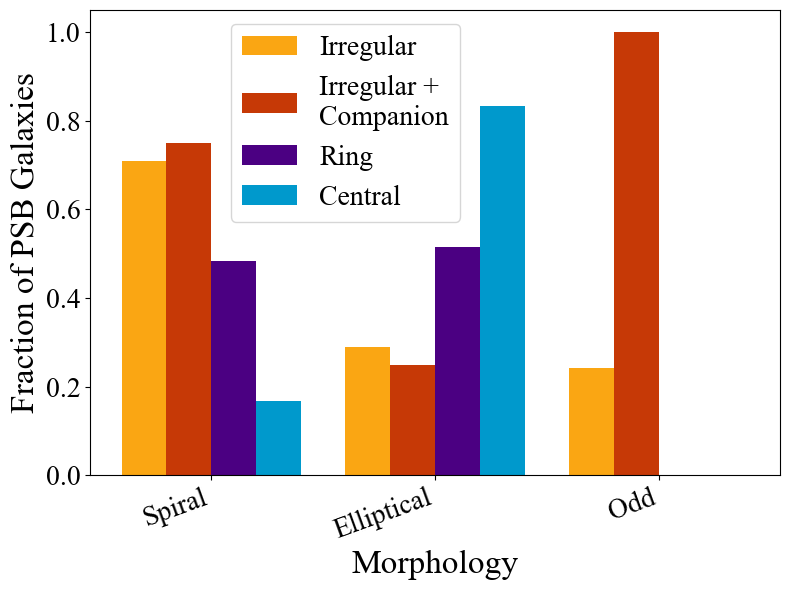}
\caption{Histogram of morphologies for PSBs selected from all three parent surveys. Odd galaxies do not represent a third category, but instead represent the fraction of both spirals and ellipticals that have odd features. The left panel compares our Summed Burst vs Percent Spaxel PSBs and the right panel compares our four PSB spaxel distributions. For galaxies meeting the Percent Spaxel Burst criteria, Irregulars, and Irregular + Companion galaxies, we see that spirals are more common than ellipticals. For galaxies meeting the Summed Burst criteria and Centrals, however, we see a preference for ellipticals.
\label{fig:hist}}
\end{figure*}

Using data from Galaxy Zoo, we search for trends in the morphologies of the galaxies in our selection. Nineteen percent of the full sample of MaNGA, 56\% of CALIFA, and 89\% of AMUSING++ do not have data in Galaxy Zoo. For PSBs without Galaxy Zoo matches, we\footnote{S. Starecheski} visually classified the morphologies using the same process as Galaxy Zoo 2. Figure \ref{fig:hist} shows the percentages of spirals vs ellipticals in our sample. The ``Odd" category in Galaxy Zoo represents the percentage of both spirals and ellipticals with odd features. A spiral with an odd feature, for example: a merger, is counted as both a ``Spiral" and an ``Odd".

We find that Summed Burst galaxies are more often ellipticals, and Percent Spaxel galaxies are more often spirals. For our Summed Burst cases, we see that majority of them are ellipticals in MaNGA (76.7\%), but this is not the case for CALIFA and AMUSING++ (42.9\% and 0\% ellipticals, respectively). We again note that the parent selections of the three IFU surveys result in different galaxy populations, which could be the cause of the different morphological distributions in our PSBs, as the CALIFA and AMUSING++ galaxies have a higher fraction of spirals than the MaNGA sample. In the full samples of all three surveys, spirals are more common than ellipticals, with CALIFA and AMUSING++ having smaller fractions of ellipticals when compared to MaNGA. We further explore why differences between surveys could lead to different results in Section \ref{betweensurveys}. For galaxies meeting our Percent Spaxel Burst criteria, we primarily see spirals in MaNGA, CALIFA, and AMUSING++ (77.4\%, 83.3\%, and 83.3\%). This is consistent with what we see for the full sample of galaxies in these surveys.  

Turning to the Right Panel of Figure \ref{fig:hist}, we see that Irregulars and Irregular + Companions are most likely to be spirals, whereas Centrals are more likely to be ellipticals. Rings are both spirals and ellipticals in approximately equal numbers. Although Rings are split between spirals and ellipticals, the tendency for the Irregulars to be spirals and the Centrals to be ellipticals seen in the right panel of Figure \ref{fig:hist} mimics the same difference we see between the Percent Spaxel Burst PSBs and the Summed Burst PSBs in the left panel. The tendency for Centrals to be ellipticals is corroborated by the preference for Centrals to fall low in measures of ongoing star formation (discussed in Section \ref{subsec:global}).

Galaxy Zoo also contains information about subclasses for the spiral, elliptical, and odd morphologies. These subclasses describe visible morphological features. By construction, 75\% of the Irregular + Companion galaxies display the odd feature ``merger". While many PSBs do not show obvious merger features they may still have experienced mergers in their recent histories, as merger features can fade on $<$ 200 Myr timescales \citep{2016MNRAS.456.3032P}.

We only see two odd galaxies with the ring morphological feature identified by Galaxy Zoo in the 3-color images. These both have an Irregular PSB spaxel distribution, indicating that the Ring distribution of PSB spaxels does not necessarily overlap with the ring morphological feature in a galaxy. Examining these two galaxies, we see large sections of the morphological ring that are star-forming and visually blue, and thus unlikely to be quenched. While in principle an outside-in or inside-out quenching pattern could appear as a concentric pattern of rings seen in the images and rings seen in the PSB spaxels, we do not see any such cases. We return to the question of quenching patterns in Section \ref{subsec:radial}. 

We do not see any barred galaxies with a Ring PSB spaxel distribution, which is a notably different result than found in \cite{2019MNRAS.489.5709C}. Allowing older bursts in our selection criteria might give time for a bar to dissipate or otherwise be disrupted. Further comparison between our results and those of \cite{2019MNRAS.489.5709C} can be found in Section \ref{subsec: betweencriteria}.

The final morphological analysis we performed involved examining the trends in the space of effective radius ($\mathrm{R}_{\mathrm{e}}$) vs stellar mass to test the compactness of our PSB galaxies. We obtained the $\mathrm{R}_{\mathrm{e}}$ values from the same files as the stellar masses described in Section \ref{subsec:global} \citep{2003MNRAS.341...33K, 2004MNRAS.351.1151B, 2004ApJ...613..898T, 2024RMxAA..60...41S}. Interestingly, we noted that our PSBs that overlap with the known SDSS E+A and PCA galaxies are very low in $\mathrm{R}_{\mathrm{e}}$, with 6/7 being below 0.5 kpc, whereas the rest of our PSB galaxies (both the Summed Burst and Percent Spaxel) cover a wider range in $\mathrm{R}_{\mathrm{e}}$ ($\sim$ -0.5 to 1.5) and are more similar to the parent sample. A KS test comparing our total PSB sample and the subset overlapping with the traditionally selected PSBs gives a p-value of 0.001, which suggests that the smaller sizes of the traditionally-selected galaxies is statistically significant. This could suggest an evolutionary track that results in compacted E+A and PCA galaxies. Other possibilities are selection effects where the strongest bursts appear more compact, or central bursts look more compact, which would be most similar to the E+As. We are cautious not to over-interpret this result because there are only 7 such galaxies.

\subsection{The Role of AGN}
\label{subsec:AGN} 

\begin{figure*}
\plottwo{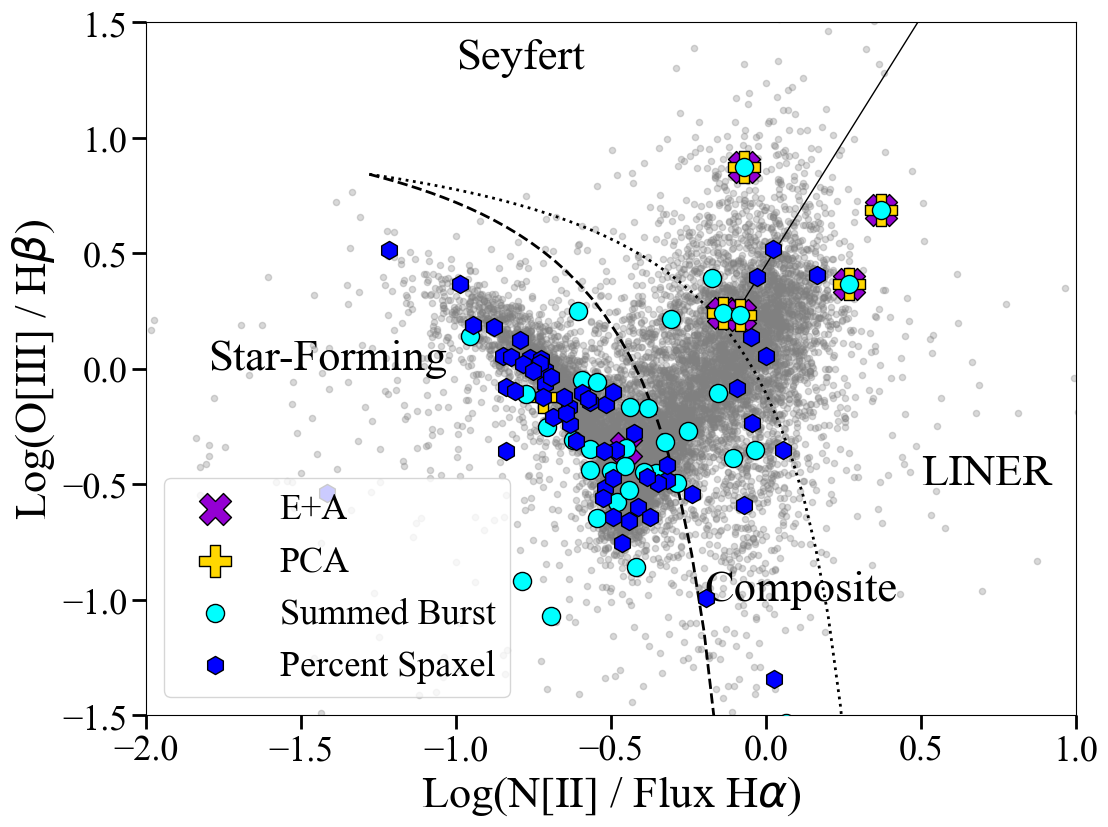}{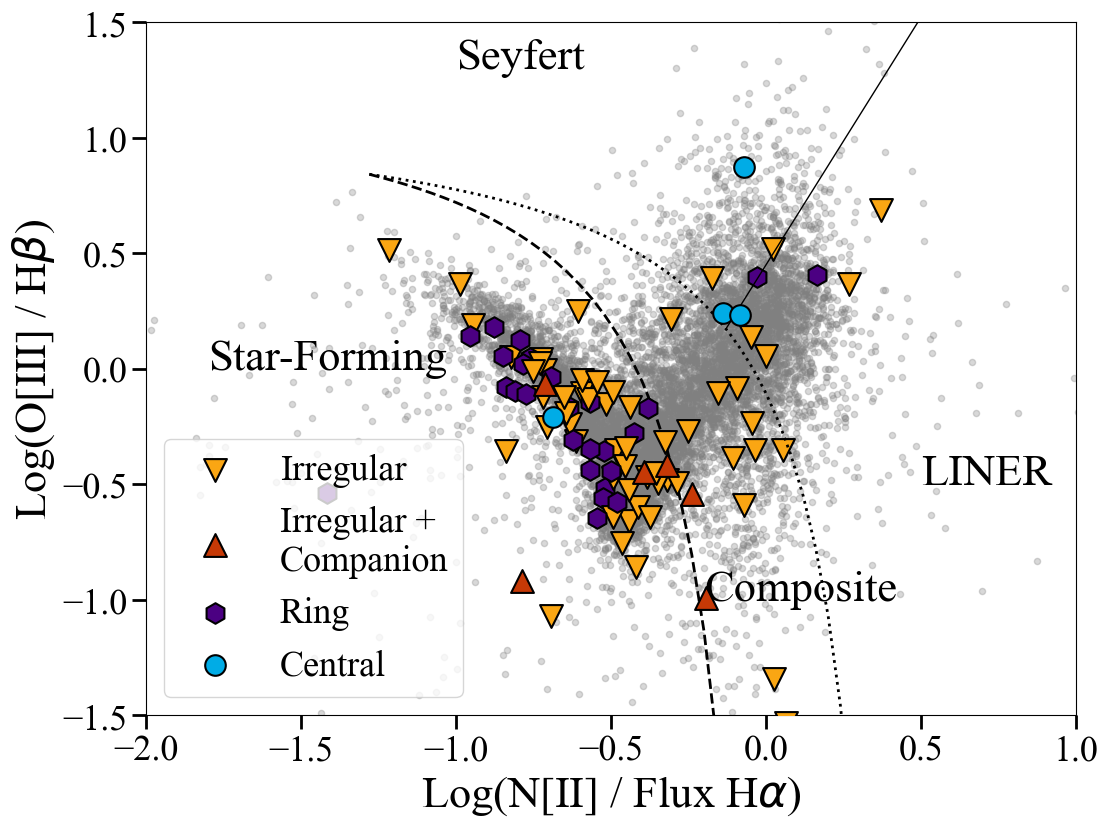}
\caption{\nii/H$\alpha$ vs \oiii/H$\beta$ BPT diagram for all galaxies in the MaNGA, CALIFA, and AMUSING++ surveys after the signal to noise cut. The dividing lines shown are from \cite{2001ApJ...556..121K} (dotted) and \cite{10.1111/j.1365-2966.2003.07154.x} (dashed and solid). In the left panel, we compare the Summed Burst PSBs vs the Percent Spaxel Burst PSBs. In the right panel, we compare our four PSB spaxel distributions. Traditional PSB selection methods are biased against strong AGN, whereas our method is not. This could have allowed us to identify a new sample of PSBs with strong Seyfert 2 lines. The absence of such a population in our sample suggests that strong AGN activity does not typically occur throughout the PSB phase.  \label{fig:bpt}}
\end{figure*}

We use optical emission line ratios to study Active Galactic Nuclei (AGN) activity during the PSB phase. Figure \ref{fig:bpt} shows the \nii/H$\alpha$ vs \oiii/ H$\beta$ BPT diagram \citep{1981PASP...93....5B} in order to better understand the source behind the ionization of the gas. The dividing lines shown are from \cite{2001ApJ...556..121K} and \cite{10.1111/j.1365-2966.2003.07154.x}. Most of our PSBs lie in the star-forming region of the plot, with scatter into the composite, LINER and Seyfert regions seen for both the Summed Burst and Percent Spaxel classifications. The galaxies in our sample that overlap with the E+A and PCA selections are mostly (5/7) located in the LINER/Seyfert region of the parameter space.

In the right panel, we see that our Rings tend to fall in the star-forming region, with many clustering on the lefthand side, in the low \nii/H$\alpha$ portion of this region. On the other hand, our Irregulars display a wider range of scatter, both across the star-forming region of the diagram, and into the other regions. This could indicate that the Irregulars may have more contribution from shocks, evolved stars, and AGN in their ionization. Half of the Centrals lie in the Seyfert and LINER regions indicating multiple sources of ionization, while the other three are star-forming or not detected.

We also consider the positions of our sample on a WHaN diagram \citep{2011MNRAS.413.1687C} as an alternate test of the ionization sources in our PSBs. Again, we see very few PSBs in the Seyfert region of the diagram, and see that most of the cases with high N[II]/H$\alpha$ have low EW H$\alpha$, which indicates that they are possibly driven by shocks or evolved stars instead of AGN.

Consistent with the results of \S\ref{subsec:global} and \S\ref{subsec:galzoo}, we find that the differences we see between the Summed Burst and Percent Spaxel Burst classifications map to differences in distributions of the PSB spaxels. We can see that the Summed Burst PSBs, especially many of those that overlap with the known SDSS PSBs, scatter into the Composite, LINER, and Seyfert regions of the plot, similar to what we see with the Centrals.

Traditional PSB selection methods are biased against strong AGN, as line emission can be confused for ongoing star formation. Our method is not subject to confusion from strong narrow line emission (although blue continuum from luminous AGN could still influence our selection). In principle, this could have allowed us to identify a new sample of PSBs with strong Seyfert 2 lines. The absence of such a population in our sample suggests that such objects are rare. This is consistent with the idea that strong AGN activity does not typically occur throughout the PSB phase \citep{10.1093/mnras/stu141, 2022ApJ...935...29L, 2025MNRAS.539.3568A, 2025ApJ...992..123F}. We consider the role of intermittent AGN activity in Section \ref{quenching}. 

\subsection{Radial Trends in Post-Burst Age}
\label{subsec:radial}
The spatially-resolved SFHs allow us to explore whether our PSB galaxies are undergoing inside-out or outside-in quenching, and whether differing radial trends drive the differences between our Ring, Central, and Irregular galaxies. In galaxies undergoing outside-in quenching, star formation would shut down in the outskirts first, showing young stellar populations in the center and older stellar populations at larger radii. Galaxies that are undergoing inside-out quenching would show the opposite. These galaxies would first suppress star formation in their centers, showing older stellar populations there, with younger stellar populations and ongoing star formation persisting at larger radii. If quenching is complete and happens slowly enough, we would see quiescent spaxels in the center and star-forming spaxels in the outskirts. But having SFH data allows us to look for trends in systems where quenching happens rapidly or systems where quenching is incomplete and star formation may resume.  

\begin{figure*}
\centering
\includegraphics[width=\textwidth]{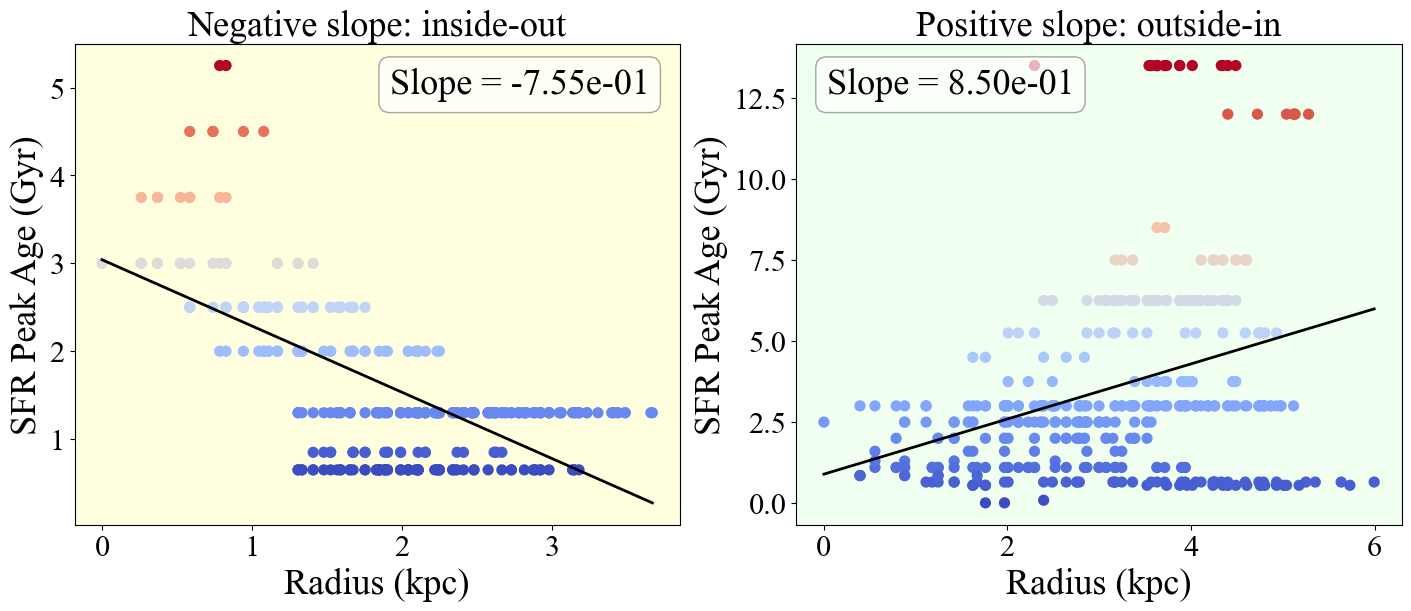}
\caption{Example of radial trends in burst age for one galaxy with a negative SFH gradient (9875-12703) and one with a positive SFH gradient (12067-3701). Each point on these plots is a spaxel in the galaxy, after the signal to noise cut was performed. The x-axis represents the radius of the galaxy in kiloparsec and the y-axis represents the age that the spaxel was at a maximum in SFR. These points are color coded by the y-axis data for readability, with red for older peak ages, and blue for younger peak ages. We have also performed a simple linear fit to the points, with the line shown in black and the line's slope displayed in the legend. Positive slopes can be read as outside-in quenching, and negative slopes can be read as inside-out quenching.}
\label{fig:slopes}
\end{figure*}

\begin{figure}
\centering
\includegraphics[width=0.5\textwidth]{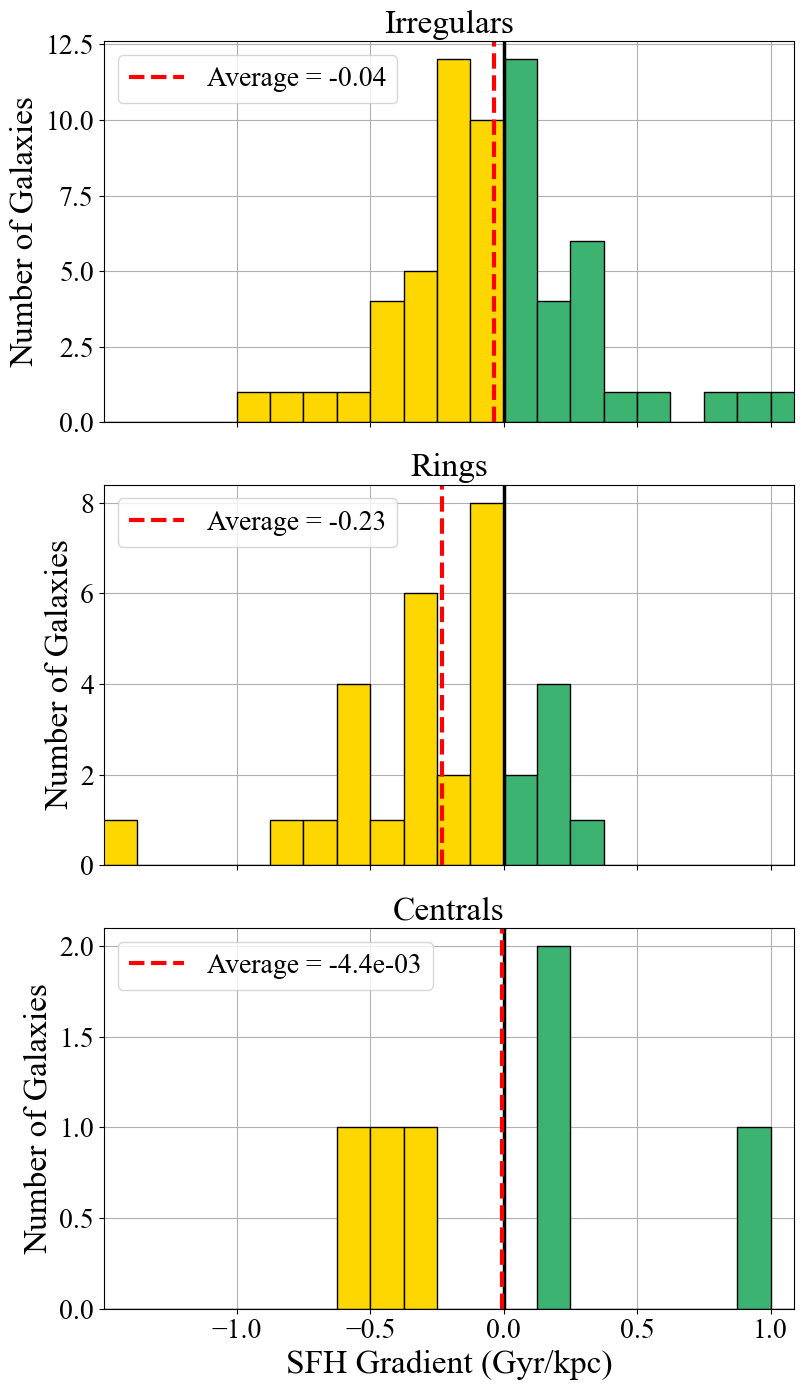}
\caption{Histogram showing the distribution of SFH gradients for our Irregular, Ring, and Central galaxies. The average for each category is plotted with a red, dashed, vertical line, and its value is listed in the legend. These ``SFH gradients" are the slope of a linear fitting to SFR vs radius data for each spaxel, with examples of this shown in Figure \ref{fig:slopes}. For easy readability, the positive SFH gradients have been colored in green, and the negative SFH gradients have been colored in yellow. Positive SFH gradients can be read as outside-in quenching, and negative SFH gradients can be read as inside-out quenching. Note that while the Irregulars and Centrals seem to show a mix of different SFH gradients, the Ring galaxies show negative SFH gradients more often than not.
\label{fig:posvneg}}
\end{figure}

We test this by considering the age at which each spaxel experiences a peak in SFR. Then we plot this age as a function of radius for each spaxel in a galaxy, and fit a straight line to this distribution. If this line has a negative slope (negative SFH gradient), the galaxy is experiencing inside-out quenching, as this indicates that the SFR peaked at older times at low radius, and peaked at recent times at high radius. Similarly, if this line has a positive slope (positive SFH gradient), the galaxy is experiencing outside-in quenching. An example of each of these is shown in Figure \ref{fig:slopes}. A histogram summarizing the distribution of SFH gradients for Irregulars, Rings, and Centrals, is shown in Figure \ref{fig:posvneg}, as well as the average SFH gradient value for each type.

For the Irregulars and Centrals, we see a mix of both positive and negative SFH gradients, where 57\% of Irregulars and 50\% of Centrals have a negative SFH gradient. Notably, for Rings, the majority of galaxies (77\%) have a negative SFH gradient, indicating that they are undergoing inside-out quenching. Inside-out quenching is often attributed to AGN activity inducing quenching in a galaxy, as the energy from the AGN works its way to the outskirts over time. A bias toward negative slopes as strong as that observed in the Ring galaxies only occurs 0.01\% of the time when randomly sampling 31 Irregular galaxies, indicating that it is highly unlikely they are drawn from the same distribution. It is difficult to tell whether or not Rings and Centrals are different stages of the same quenching process, as we have relatively few Centrals in our sample. Both our methods and results for assessing radial trends in quenching patterns differ from previous works in the literature, which we discuss further in \S\ref{quenching}. 

We consider whether environment density could play a role in producing these different SFH gradients. 
We consider the density of galaxies within the radius to the 5th nearest neighbor, using the catalogs by \cite{2015MNRAS.451..660E} for MaNGA and \cite{2017ApJ...848...87C} for CALIFA. We find no statistically significant difference between the environment densities of our positive and negative SFH gradient galaxies, with a KS test yielding a p-value of 0.7. It should be noted that this analysis was performed on a subset of our full PSB sample, as density measurements were only available for 17 of our MaNGA PSBs and 2 of our CALIFA PSBs. Therefore, trends between quenching pattern and environment density may emerge if a more complete dataset becomes available.

\begin{figure}
\centering
\includegraphics[width=0.5\textwidth]{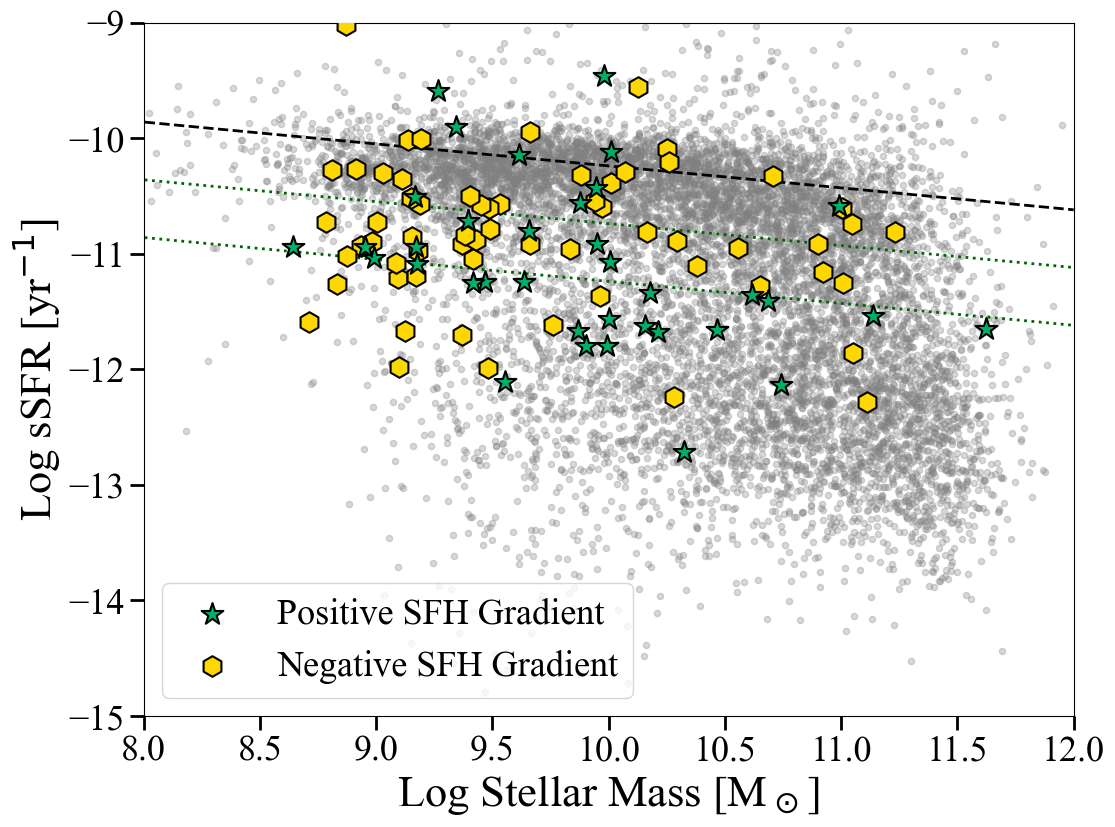}
\caption{Stellar mass vs sSFR for all galaxies in the MaNGA, CALIFA, and AMUSING++ surveys after the signal to noise cut. Lines indicating the main sequence and green valley are drawn in the same manner as Figure \ref{fig:stellarmasssfr}.  PSBs with a positive SFH gradient (see \S4.2) tend to have lower sSFR values than PSBs with a negative SFH gradient, which is interesting since we expect the highly-weighted star-forming center to dominate over the quenched outskirts.
\label{fig:stellarmasssfr3}}
\end{figure}

Finally, we examine where our PSB galaxies with positive vs negative SFH gradients fall on various diagnostic plots. On a specific SFR vs stellar mass plot (shown in Figure \ref{fig:stellarmasssfr3}), galaxies with a positive SFH gradient tend to have lower sSFR values than galaxies with a negative SFH gradient. A KS test between these two distributions yields a p-value of 0.02, indicating a reasonable level of significance. Because the center will be more highly weighted when we take the light-weighted average, it is unusual that galaxies quenching outside-in have lower average sSFR values, since we expect the highly-weighted star-forming center to dominate over the quenched outskirts. For the outside-in galaxies with low SFRs, this indicates that while the peak SFR was 1 Gyr ago, it has since quenched. Meanwhile, in the inside-out galaxies with high SFRs, this indicates that while the bulk of the stars in the center were formed at least 3-5 Gyr ago, star formation has resumed in the center. A KS test for stellar mass yields a p-value of 0.11, suggesting that these two samples could have been from the same parent sample and the difference between them is not statistically significant along this axis. We also examined the Galaxy Zoo morphologies of these galaxies to see if there were any differences between those with positive and those with negative SFH gradients. We found that the positive SFH gradient galaxies are fairly evenly distributed between spirals and ellipticals, with around a 50/50 split, (51.4\% and 48.7\%, respectively), but that the negative SFH gradient galaxies are more likely to be spirals than ellipticals, (66.1\% and 33.9\%). The tendency for galaxies with positive SFH gradients to be more frequently quenched than those with negative gradients aligns with the trends observed in the specific SFR vs stellar mass plot. We also see that negative SFH gradient PSBs are more likely to have bars than the positive SFH gradient PSBs (12 vs 2). Finally, on a \nii/H$\alpha$ vs \oiii/ H$\beta$ BPT diagram, positive and negative SFH gradient galaxies are both approximately equally likely to appear in the AGN region of the diagram, with a KS test revealing no significant evidence that these positive and negative SFH gradient galaxies come from different distributions. We find no indication that inside-out galaxies preferentially occupy the AGN region of the BPT diagram compared to outside-in galaxies. The implications of this are discussed further in section \S\ref{quenching}.

\section{Discussion} \label{sec:discussion}
\subsection{Comparison Between Surveys}
\label{betweensurveys}

We see several places where the MaNGA and CALIFA parent samples show differing global properties for the SFH selected PSB samples. We avoid drawing conclusions about systematic differences between AMUSING++ and the other two surveys because of the small number of SFH selected PSBs in AMUSING++. While the star-forming galaxies selected in MaNGA and CALIFA are similar, the quiescent galaxies selected from MaNGA are typically lower mass than those selected by CALIFA. The quiescent PSBs selected from MaNGA also typically show higher values of Lick H$\delta_A$. 

These differences could be because MaNGA selects galaxies out to a higher redshift, and thus probes a larger volume than CALIFA, making it easier to identify rare events like enhanced Balmer absorption from bursts. However, CALIFA galaxies are observed to a smaller physical resolution, so the CALIFA galaxies may be more strongly influenced by smaller scale features. The CALIFA galaxies are also observed out to a larger physical extent than the MaNGA galaxies. If there are PSB features that are small in scale or PSB regions further out in the outskirts of otherwise star-forming galaxies, CALIFA is more likely to see them, whereas in MaNGA they would be washed out and blend in with the rest of the galaxy, or would be outside the field of view. These differences could also be observed due to inherent differences in the galaxies sampled, or the result of different targeting across surveys (such as AMUSING++ having a bias towards spirals because of supernova follow up). 

\subsection{Comparison with Other PSB Selections}
\label{subsec: betweencriteria}   

Several recent works in the literature have used IFU data to select and characterize PSB galaxies, though the methods and findings vary \citep{2018MNRAS.480.2544R, 2019MNRAS.489.5709C, 2024ApJ...961..216C, 2024MNRAS.528.4029L, 2025A&A...696A.206V}. Ours is complementary to these studies; we are selecting PSBs directly from the SFHs of these galaxies, but can still compare our samples and their properties. 

Despite differences in selection methods, several of the galaxies we select here are also identified by the other searches referenced above. We find that 8 of our MaNGA PSB galaxies (all from our Summed Burst selection criteria) overlap with the sample of 48 PSBs outlined in \cite{2024MNRAS.528.4029L}. Turning to \cite{2025A&A...696A.206V}, we find that 2 of their 7 PSBs overlap with our PSB sample. Finally, \cite{2019MNRAS.489.5709C} report coordinates for their Rings and Centrals only, from which we identify 2 Rings and 4 Centrals (out of 68 Rings and Centrals). For all six overlapping galaxies, our spaxel morphology classifications differ from those reported in \cite{2019MNRAS.489.5709C}, at least in part because of the stricter Ring criteria we adopted.

Compared to other studies, we find a higher percentage of Rings (29\%) and fewer Centrals (5.6\%). \cite{2024ApJ...961..216C} find 310 Irregular galaxies (63\%), 85 Ring galaxies (17\%), and 94 Central galaxies (19\%), for a total of 489 PSBs. \cite{2019MNRAS.489.5709C} examine a total of 360 galaxies with PSB regions: 31 Centrals (8.6\%), 37 Rings (10.3\%), and 292 Irregular galaxies (81.1\%).

The Galaxy Zoo morphologies of our selected PSBs show similar morphological trends as those exhibited by the PSBs in \cite{2025A&A...696A.206V}. \cite{2025A&A...696A.206V} find that in general, their PSB galaxies do not have a disc with spiral arms, but are in transition to an elliptical galaxy. In the left panel of Figure \ref{fig:hist}, we see that this is often true of our Summed Burst PSBs, which are most analogous to the selection method used in \cite{2025A&A...696A.206V}. They also find that 2 out of these 7 (29\%) are lenticular galaxies. A similar fraction (22$\%$) of the galaxies we classified by eye are lenticular galaxies.

In contrast to \cite{2019MNRAS.489.5709C}, we do not find an excess of bar features in the galaxies we select as Ring PSBs. Although 19\% of Ring galaxies show signs of a bar in \cite{2019MNRAS.489.5709C}, none of our Rings are classified as having bars by Galaxy Zoo. Additionally, only $2/31 = 6.5\%$ have bars listed in their morphologies in HyperLeda\footnote{http://leda.univ-lyon1.fr/} \citep{2014A&A...570A..13M}, and even then, these are only classified as an ``intermediate bar" and are not found in every dataset for these objects. The differences in properties of our Ring galaxies compared to the Ring galaxies of \cite{2019MNRAS.489.5709C} are most likely due to the stricter criteria we required for a galaxy to be classified as a Ring PSB. 

Our PSBs tend to lie primarily below the average main sequence line. This result is consistent with \cite{2025A&A...696A.206V}, who tend to see their PSBs on the lower region of the main sequence, green valley, and quenched region (as opposed to majority of their post-merger galaxies which tend to lie on the star forming main sequence). On the other hand, \citet{2024ApJ...961..216C} find their PSBs to be mostly star-forming or in the green valley (rather than quenched). \cite{2025A&A...696A.206V} find that their PSB galaxies that have a central concentration of PSB spaxels often have higher integrated SFRs, as opposed to their PSB galaxies with spatially distributed PSB emission. This is in contrast to what we see in Figure \ref{fig:stellarmasssfr}, where our Summed Burst PSBs (analogous to our Centrals), fall lower in terms of specific SFR than our Percent Spaxel Burst PSBs. 

We find similar results examining the relative star forming properties for the 4 different PSB spaxel morphologies. \citet{2024ApJ...961..216C} find their Centrals to be in the green valley more often than not (48 out of 94), and find that quite often their Centrals have significant star-forming regions spatially. In contrast with this, in the right panel of Figure \ref{fig:stellarmasssfr}, we see that our Centrals are more often globally quenched (4/6), although we do have a relatively small fraction of Centrals. However, most of our Centrals are from our Summed Burst criteria, meaning that although the highly-weighted galactic centers are quenched, ongoing star formation could occur in spatially large regions of the outskirts of these galaxies. Furthermore, \citet{2024ApJ...961..216C} find that 82 out of 85 Ring galaxies are star-forming. This is in contrast to our finding that many of our Rings lie in the green valley, with the caveat that our definitions of the green valley are not identical. 

\begin{figure*}
\plottwo{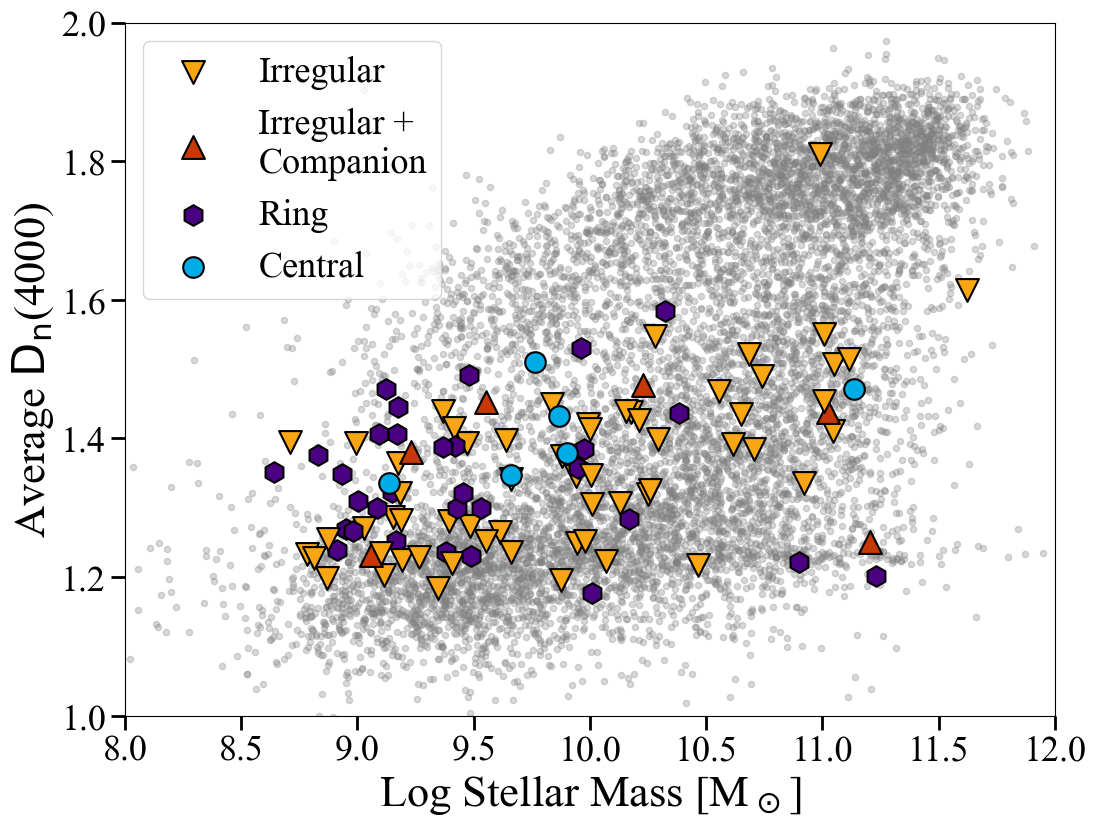}{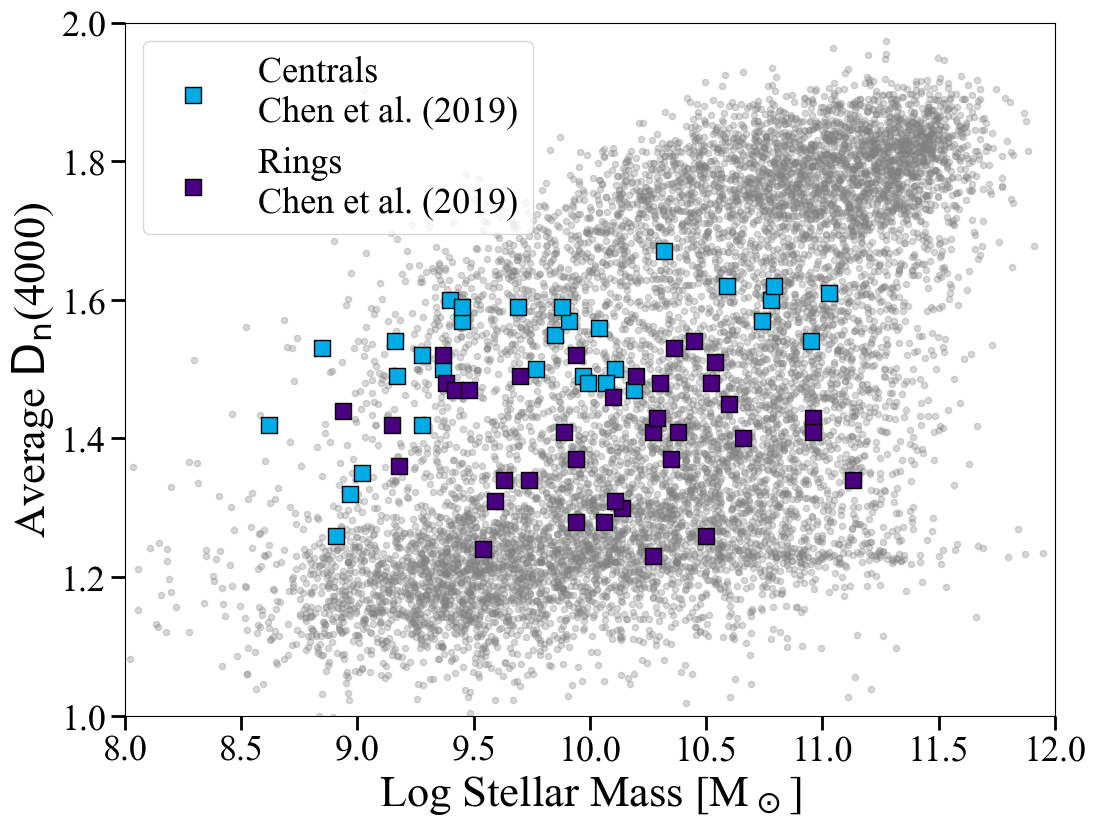}
\caption{Logarithm of stellar mass vs average \dfour index for all galaxies in the MaNGA, CALIFA, and AMUSING++ surveys after the signal to noise cut. Modeled after Figure 5 in \cite{2019MNRAS.489.5709C}, with our objects shown on the left, and the Rings and Centrals from \cite{2019MNRAS.489.5709C} shown on the right. Although it is difficult to draw conclusions because of the relatively small number of Centrals we observe in our sample, they tend to lie in the green valley. However, we do not see the same clear division between Rings and Centrals as \cite{2019MNRAS.489.5709C}, and instead see significant overlap between the Centrals and other PSB morphological types. 
\label{fig:d400}}
\end{figure*}

\citet{2019MNRAS.489.5709C} use \dfour to trace the mean stellar ages of their galaxies, finding that their Centrals have systematically higher \dfour (and thus older mean stellar ages) than their Rings. In Figure \ref{fig:d400}, we compare our division of PSB morphologies to that of \citet{2019MNRAS.489.5709C}. We do not see the same clear division, and instead see significant overlap between the Centrals and other PSB morphological types. Although we do not observe any Centrals in the lower region of the diagram, it is difficult to draw conclusions because of the relatively small number of Centrals we find in our sample. We also consider \dfour values measured at the effective radius instead of the light-weighted average \dfour for our MaNGA galaxies, and found that this change did not alter our conclusions. 

To summarize, we find a small amount of overlap with other PSB samples in an object-by-object comparison, reinforcing the idea that we are presenting a unique selection of PSBs. We generally find a smaller percentage of Central galaxies and a higher percentage of Ring galaxies compared to other studies. We find broad agreement between our Galaxy Zoo morphologies and morphological trends reported in other studies, although our Ring galaxies do not exhibit the bars reported elsewhere. In Figure \ref{fig:stellarmasssfr}, we generally see that our Summed Burst PSBs/Centrals fall lower in terms of sSFR than in other PSB studies. In Figure \ref{fig:d400}, we do not see the same clear division between Centrals and Rings reported in \citet{2019MNRAS.489.5709C}.

\subsection{Implications for Quenching Pathways}
\label{quenching}

From Figure \ref{fig:stellarmasssfr}, we can see that many of our PSB galaxies have ongoing star formation outside of the PSB regions, and may not be globally quenching. It can be difficult to discern whether these star-forming regions are on the verge of quenching, or whether the PSB regions are about to reignite. Particularly for the Ring galaxies that lie on the main sequence of star forming galaxies, there may have been only a ring undergoing a burst of star formation that subsequently quenched, while star formation continues elsewhere in the galaxy, suggesting that in many cases the PSB properties we observe are not indicative of a trend for the entire galaxy. Galaxies may take different paths to quiescence and further research will be needed to understand how to differentiate between these pathways. 

Mergers may be one way of inciting quenching that would produce unique signatures in terms of where the PSB regions are, although the role of mergers in driving starbursts is subject to debate \citep{2003A&A...405...31B, 2023ApJ...952..122G}. \cite{2018MNRAS.480.2544R} show that in contrast to simulations of mergers, which predict the funneling of gas towards the center of the galaxy leading to central starbursts, their observations find an enhancement of star formation in the outskirts of the galaxies. On the other hand, \cite{2024A&A...690L...4T} look at radial trends in SFR for 150 post-merger galaxies in MaNGA to see if starbursts triggered by mergers are discernible from secular starbursts, and find that mergers trigger a much stronger central starburst. \cite{2018MNRAS.474.2039E}, also using MaNGA data, create a new metric -- $\Delta\Sigma_{\mathrm{SFR}}$ -- to look at spaxel-resolved radial SFR profiles. Although they do see a central enhancement of star-formation in galaxies engaged in mergers, they find no significant change in their results if they exclude spaxels in merging galaxies from their sample, suggesting that other processes may drive this boost in central SFR. Our Irregular + Companion galaxies, which are undergoing mergers, have a majority of their PSB regions on the outskirts of the galaxy in Appendix \ref{app:1}, indicating a starburst occurred on the outskirts in the recent past. This is in line with the observations of \cite{2018MNRAS.480.2544R} and in contrast with the findings of \cite{2024A&A...690L...4T}. These findings may indicate that the starburst and quenching locations vary between the pre-coalescence and post-merger phases. 

Previous studies of Ring and Central PSBs have disagreed on whether these objects represent multiple pathways or multiple stages of evolution. \cite{2019MNRAS.489.5709C} conclude that Rings and Centrals are not different evolutionary phases of the same quenching process, suggesting that their different quenching pathways are most likely the result of different physical mechanisms, where Centrals occur after a major disruptive event and Rings occur due to an interference in gas fueling to the outer regions of a galaxy. On the other hand, \cite{2024ApJ...961..216C} posits that Rings and Centrals are two different stages of outside-in quenching, whereas they find their Irregular galaxies to follow inside-out quenching. \cite{2025A&A...696A.206V}, which consider galaxies analogous to our Centrals, find evidence that their PSBs are undergoing an outside-in quenching process. They propose that the different locations of their PSB galaxies on a SFR vs stellar mass plot represent different stages of the same quenching process, which can be differentiated between because these processes occur slowly in an isolated environment after a merger. 

In contrast to \cite{2024ApJ...961..216C}, in Figure \ref{fig:posvneg}, we do not see strong evidence for exclusively outside-in quenching in our Rings, Centrals, or Irregulars, suggesting that inside-out or outside-in might be too simplistic a model for many galaxies. However, we do see that our Ring galaxies are more often than not following inside-out quenching. This difference from previous work is most likely due to differences in how we look for inside-out and outside-in quenching. By examining the SFH of these galaxies, we are looking directly at the ages that star formation occurs and can separate out new from old star formation. \cite{2024ApJ...961..216C} on the other hand looks for gradients in \dfour, H$\alpha$, and H$\delta_A$, meaning their observations are sensitive to the relative fractions of young and old stellar populations in addition to age. Ultimately, it is difficult to tell in our sample whether or not Rings and Centrals are different stages of the same quenching process, as we have relatively few Centrals in our sample. 

As discussed in \S\ref{subsec:AGN}, our methods do not suffer from the same biases against strong AGN that affect traditional PSB selection techniques. This could have allowed us to identify a new population of PSBs with strong Seyfert 2 lines. Their absence in our sample reinforces the idea that these objects are rare. This is consistent with the idea that strong AGN activity does not typically occur throughout the PSB phase \citep{10.1093/mnras/stu141, 2022ApJ...935...29L, 2025MNRAS.539.3568A, 2025ApJ...992..123F}, which we also see more generally for our entire PSB sample in Figure \ref{fig:bpt}. Despite not seeing evidence for this on a BPT diagram, if the AGN duty cycles occur quickly enough, they will wash out any trends with the current AGN luminosity. 

Inside-out quenching is usually attributed to AGN activity prompting quenching within a galaxy, as the energy from the AGN works its way to the outskirts over time. Therefore, if there was AGN induced quenching occurring in our sample, we would expect our inside-out PSB galaxies to be more common among galaxies with AGN-like emission line ratios compared with the outside-in PSBs. In \S\ref{subsec:radial}, we find no evidence that inside-out galaxies preferentially occupy the AGN region of the BPT diagram. Thus, if AGN feedback is the cause for the inside-out quenching pattern we see, particularly for the Ring galaxies, it must occur with AGN duty cycles much shorter than the post-starburst phase. 

While bars have been proposed as a key secular method of quenching galaxies, particularly in Ring PSBs \citep{2019MNRAS.489.5709C, 2019MNRAS.490.2347Q}, we do not see any evidence of this effect in our Ring PSBs, as none are classified as having bars in Galaxy Zoo. However, given that bars provide another potential route to inside-out quenching, it is notable that we do identify 12 barred galaxies in our negative SFH gradient PSBs, while only seeing 2 in our positive SFH gradient PSBs. Although \cite{2019MNRAS.490.2347Q} underscore the challenges of finding bars in the SDSS optical images of their 10 recently quenching MaNGA galaxies, they still suggest the gravitational potential of galactic bars as a possible cause of the (ring-like) quenching patterns that they see, citing the existence of hidden bars in spiral galaxies. A similar caveat may apply to our findings. 

It is possible that the evolutionary paths of PSBs vary as a function of stellar mass, and \cite{2018MNRAS.480.2544R} show that high mass galaxies follow a pattern of inside-out growth and quenching, in contrast to low mass galaxies that are building mass at all radii. As discussed in \S\ref{subsec:radial}, we do not see any trends in stellar mass in Figure \ref{fig:stellarmasssfr3}, and a KS test between these two samples suggests that they are indistinguishable in terms of stellar mass. Due to the lack of very many high mass galaxies in our sample, it is ultimately difficult to directly compare to \cite{2018MNRAS.480.2544R}. However, their notion that lower mass galaxies are building mass at all radii could explain the mix of SFH gradients we see in our Irregulars and Centrals in Figure \ref{fig:posvneg}.

Overall, it is difficult to draw any conclusions as to why the Rings are preferentially undergoing inside-out quenching and this remains an open-ended question that requires further study. 

\section{Conclusion} \label{sec:conclusion}

We have developed a set of criteria to select PSB galaxies based on star formation history measurements from Pipe3D, finding a total of 107 PSBs in the MaNGA, CALIFA, and AMUSING++ surveys. Our SFH selection is calibrated on E+A galaxies, and requires that $<0.13$\% of the galaxy's mass was formed in the most recent 100 Myr, while $>4.8$\% of the mass was formed in at least two consecutive age bins between 100 Myr -- 1.5 Gyr. We select two populations of PSBs: Summed Burst galaxies meet our criteria based on the light-weighted average of all spaxels in the galaxy; Percent Spaxel Burst galaxies contain at least 24\% spaxels meeting our criteria. 

This provides us with a new sample of PSB galaxies that can be used to learn about their properties as a population and probe the processes that drive quenching. These methods can be used to select PSB galaxies in future IFU surveys, including surveys like AMUSING++, where Lick H$\delta_A$ is outside of the wavelength range of the MUSE instrument, causing populations selected by traditional methods such as the E+A criteria to miss PSB galaxies in these surveys. The results of our analysis of SFH selected PSB galaxies can be summarized as follows:

\begin{enumerate}

\item Our PSB selection is a new population that has not previously been studied. Compared to previous samples of PSBs selected using single-fiber data, we select 24\% of the E+A galaxies, 8\% of the PCA-selected galaxies, and no SPOGs. Of our 107 PSB galaxies, only 7 were previously-identified as PSBs. 

\item We observe differences between our Summed Burst and Percent Spaxel Burst galaxies. Our Summed Burst galaxies are more often quenched, whereas our Percent Spaxel Burst galaxies are frequently still forming stars. Many of our Percent Spaxel galaxies have ongoing star formation in areas of the galaxy outside of the PSB regions, indicating that they are not globally quenching and that galaxies may take different paths to quiescence. 

\item We examine the spatial distributions of the spaxels meeting our PSB criteria, and classify them as either Irregulars, Rings, Centrals, or Irregular + Companions. We observe a correlation in properties between our two sets of selection criteria and our PSB spaxel distributions. Specifically, we see a strong correlation between our Summed Burst PSBs and our Centrals. Because the Summed Burst criteria is based on light-weighted averages, these galaxies are more likely to be centrally post-starburst. We do not observe the high fraction of bars in galaxies with a Ring PSB spaxel distribution noted in previous studies. 

\item Traditional PSB selection  methods are biased against AGN, as it can often be difficult to distinguish ongoing star formation from AGN activity. In contrast, our approach avoids these biases. Despite this, we still find no evidence for a significant PSB population with strong Seyfert 2 characteristics. This suggests that strong AGN activity is uncommon throughout the PSB phase. 

\item We see evidence that our Ring galaxies are undergoing inside-out quenching, in contrast to what has been observed in the literature. We do not see strong evidence for preferentially inside-out or outside-in quenching in our Irregulars or Centrals. It is difficult to tell whether or not Rings and Centrals are two different stages of the same quenching process, as suggested by previous works, due to the relative scarcity of Centrals in our sample.

\end{enumerate}

\section{Acknowledgments}
We thank the referee for their detailed comments, which have improved the clarity of this paper.

This project was made possible by NSF grant \# NSF AAG 23-07440 and UIUC’s IDEAS program. We thank Alberto Bolatto for useful discussions. 

SFS thanks the support by UNAM PASPA – DGAPA. Authors acknowledges financial support from the Spanish Ministry of Science and Innovation (MICINN), project PID2019-107408GB-C43 (ESTALLIDOS).

V.V. acknowledges support from the Comité ESO Mixto 2024 and from the ANID BASAL project FB210003.

This study uses data provided by the Calar Alto Legacy Integral Field Area (CALIFA) survey (http://califa.caha.es/).

Based on observations collected at the Centro Astronómico Hispano Alemán (CAHA) at Calar Alto, operated jointly by the Max-Planck-Institut fűr Astronomie and the Instituto de Astrofísica de Andalucía (CSIC).

Funding for the Sloan Digital Sky Survey IV has been provided by the Alfred P. Sloan Foundation, the U.S. Department of Energy Office of Science, and the Participating Institutions. SDSS acknowledges support and resources from the Center for High-Performance Computing at the University of Utah. The SDSS web site is www.sdss4.org.

SDSS is managed by the Astrophysical Research Consortium for the Participating Institutions of the SDSS Collaboration including the Brazilian Participation Group, the Carnegie Institution for Science, Carnegie Mellon University, Center for Astrophysics | Harvard \& Smithsonian (CfA), the Chilean Participation Group, the French Participation Group, Instituto de Astrofísica de Canarias, The Johns Hopkins University, Kavli Institute for the Physics and Mathematics of the Universe (IPMU) / University of Tokyo, the Korean Participation Group, Lawrence Berkeley National Laboratory, Leibniz Institut für Astrophysik Potsdam (AIP), Max-Planck-Institut für Astronomie (MPIA Heidelberg), Max-Planck-Institut für Astrophysik (MPA Garching), Max-Planck-Institut für Extraterrestrische Physik (MPE), National Astronomical Observatories of China, New Mexico State University, New York University, University of Notre Dame, Observatório Nacional / MCTI, The Ohio State University, Pennsylvania State University, Shanghai Astronomical Observatory, United Kingdom Participation Group, Universidad Nacional Autónoma de México, University of Arizona, University of Colorado Boulder, University of Oxford, University of Portsmouth, University of Utah, University of Virginia, University of Washington, University of Wisconsin, Vanderbilt University, and Yale University.

The Legacy Surveys consist of three individual and complementary projects: the Dark Energy Camera Legacy Survey (DECaLS; Proposal ID \#2014B-0404; PIs: David Schlegel and Arjun Dey), the Beijing-Arizona Sky Survey (BASS; NOAO Prop. ID \#2015A-0801; PIs: Zhou Xu and Xiaohui Fan), and the Mayall z-band Legacy Survey (MzLS; Prop. ID \#2016A-0453; PI: Arjun Dey). DECaLS, BASS and MzLS together include data obtained, respectively, at the Blanco telescope, Cerro Tololo Inter-American Observatory, NSF’s NOIRLab; the Bok telescope, Steward Observatory, University of Arizona; and the Mayall telescope, Kitt Peak National Observatory, NOIRLab. Pipeline processing and analyses of the data were supported by NOIRLab and the Lawrence Berkeley National Laboratory (LBNL). The Legacy Surveys project is honored to be permitted to conduct astronomical research on Iolkam Du’ag (Kitt Peak), a mountain with particular significance to the Tohono O’odham Nation.

NOIRLab is operated by the Association of Universities for Research in Astronomy (AURA) under a cooperative agreement with the National Science Foundation. LBNL is managed by the Regents of the University of California under contract to the U.S. Department of Energy.

This project used data obtained with the Dark Energy Camera (DECam), which was constructed by the Dark Energy Survey (DES) collaboration. Funding for the DES Projects has been provided by the U.S. Department of Energy, the U.S. National Science Foundation, the Ministry of Science and Education of Spain, the Science and Technology Facilities Council of the United Kingdom, the Higher Education Funding Council for England, the National Center for Supercomputing Applications at the University of Illinois at Urbana-Champaign, the Kavli Institute of Cosmological Physics at the University of Chicago, Center for Cosmology and Astro-Particle Physics at the Ohio State University, the Mitchell Institute for Fundamental Physics and Astronomy at Texas A\&M University, Financiadora de Estudos e Projetos, Fundacao Carlos Chagas Filho de Amparo, Financiadora de Estudos e Projetos, Fundacao Carlos Chagas Filho de Amparo a Pesquisa do Estado do Rio de Janeiro, Conselho Nacional de Desenvolvimento Cientifico e Tecnologico and the Ministerio da Ciencia, Tecnologia e Inovacao, the Deutsche Forschungsgemeinschaft and the Collaborating Institutions in the Dark Energy Survey. The Collaborating Institutions are Argonne National Laboratory, the University of California at Santa Cruz, the University of Cambridge, Centro de Investigaciones Energeticas, Medioambientales y Tecnologicas-Madrid, the University of Chicago, University College London, the DES-Brazil Consortium, the University of Edinburgh, the Eidgenossische Technische Hochschule (ETH) Zurich, Fermi National Accelerator Laboratory, the University of Illinois at Urbana-Champaign, the Institut de Ciencies de l’Espai (IEEC/CSIC), the Institut de Fisica d’Altes Energies, Lawrence Berkeley National Laboratory, the Ludwig Maximilians Universitat Munchen and the associated Excellence Cluster Universe, the University of Michigan, NSF’s NOIRLab, the University of Nottingham, the Ohio State University, the University of Pennsylvania, the University of Portsmouth, SLAC National Accelerator Laboratory, Stanford University, the University of Sussex, and Texas A\&M University.

BASS is a key project of the Telescope Access Program (TAP), which has been funded by the National Astronomical Observatories of China, the Chinese Academy of Sciences (the Strategic Priority Research Program “The Emergence of Cosmological Structures” Grant \# XDB09000000), and the Special Fund for Astronomy from the Ministry of Finance. The BASS is also supported by the External Cooperation Program of Chinese Academy of Sciences (Grant \# 114A11KYSB20160057), and Chinese National Natural Science Foundation (Grant \# 12120101003, \# 11433005).


The Legacy Surveys imaging of the DESI footprint is supported by the Director, Office of Science, Office of High Energy Physics of the U.S. Department of Energy under Contract No. DE-AC02-05CH1123, by the National Energy Research Scientific Computing Center, a DOE Office of Science User Facility under the same contract; and by the U.S. National Science Foundation, Division of Astronomical Sciences under Contract No. AST-0950945 to NOAO.

The Digitized Sky Surveys were produced at the Space Telescope Science Institute under U.S. Government grant NAG W-2166. The images of these surveys are based on photographic data obtained using the Oschin Schmidt Telescope on Palomar Mountain and the UK Schmidt Telescope. The plates were processed into the present compressed digital form with the permission of these institutions.

The Second Palomar Observatory Sky Survey (POSS-II) was made by the California Institute of Technology with funds from the National Science Foundation, the National Geographic Society, the Sloan Foundation, the Samuel Oschin Foundation, and the Eastman Kodak Corporation. The Oschin Schmidt Telescope is operated by the California Institute of Technology and Palomar Observatory.

The UK Schmidt Telescope was operated by the Royal Observatory Edinburgh, with funding from the UK Science and Engineering Research Council (later the UK Particle Physics and Astronomy Research Council), until 1988 June, and thereafter by the Anglo-Australian Observatory. The blue plates of the southern Sky Atlas and its Equatorial Extension (together known as the SERC-J), as well as the Equatorial Red (ER), and the Second Epoch [red] Survey (SES) were all taken with the UK Schmidt telescope at the AAO.

We acknowledge the usage of the HyperLeda database (http://leda.univ-lyon1.fr). 

Software: {\tt Astropy} \citep{2022ApJ...935..167A}.

\appendix
\section{PSB Spaxel Distribution Maps}
\label{app:1}

In Figure \ref{fig:maps}, we present maps of the PSB spaxel distributions. PSB spaxels and non-PSB spaxels are shown in green and blue, respectively.  Parenthesis containing an M, C, or A are given after the galaxy identifiers, designating whether the parent survey was MaNGA, CALIFA, or AMUSING++, respectively.

\begin{figure}
\centering
\includegraphics[width=\textwidth]{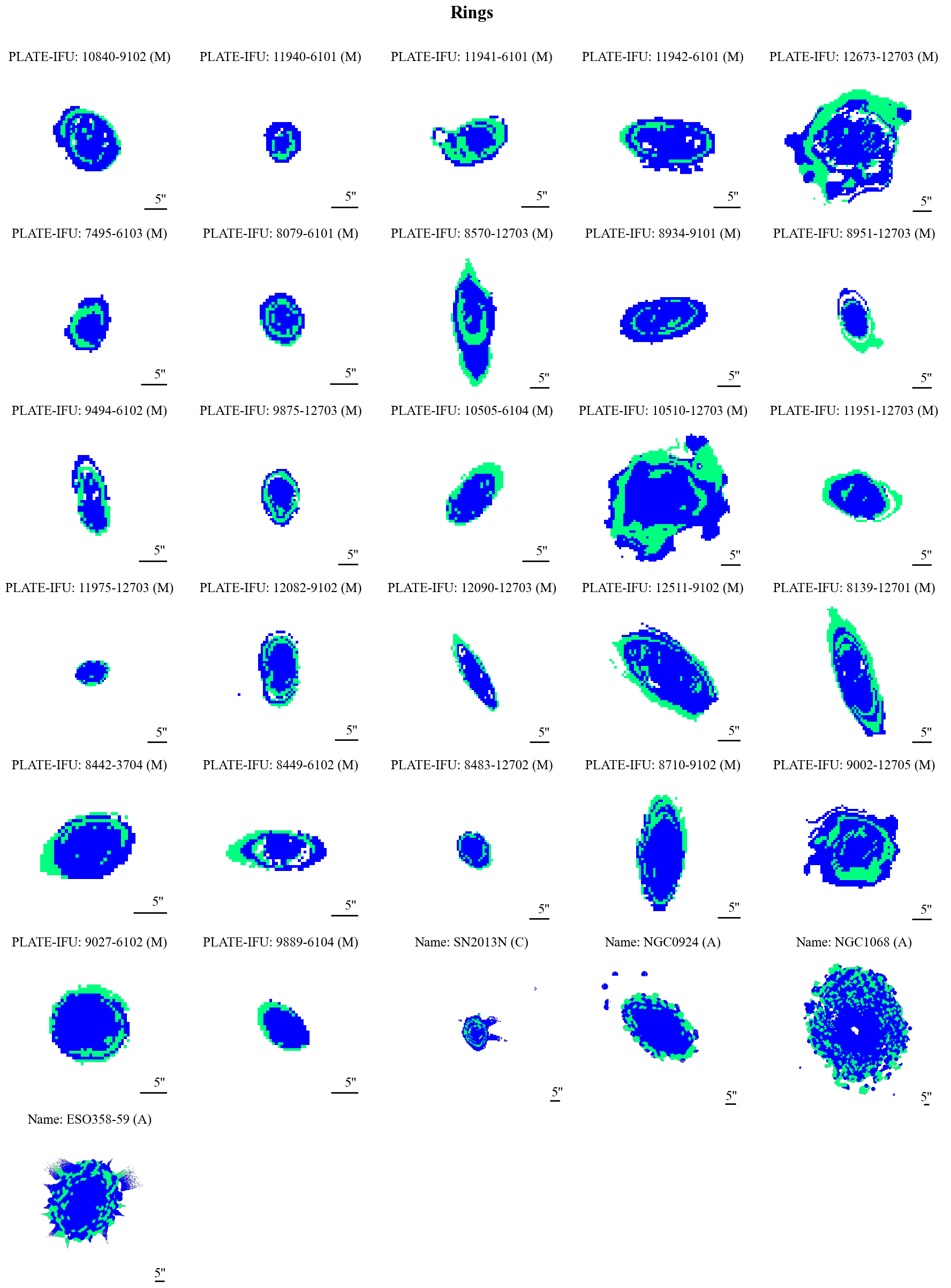}
\end{figure}
\clearpage

\begin{figure}
\centering
\includegraphics[width=\textwidth]{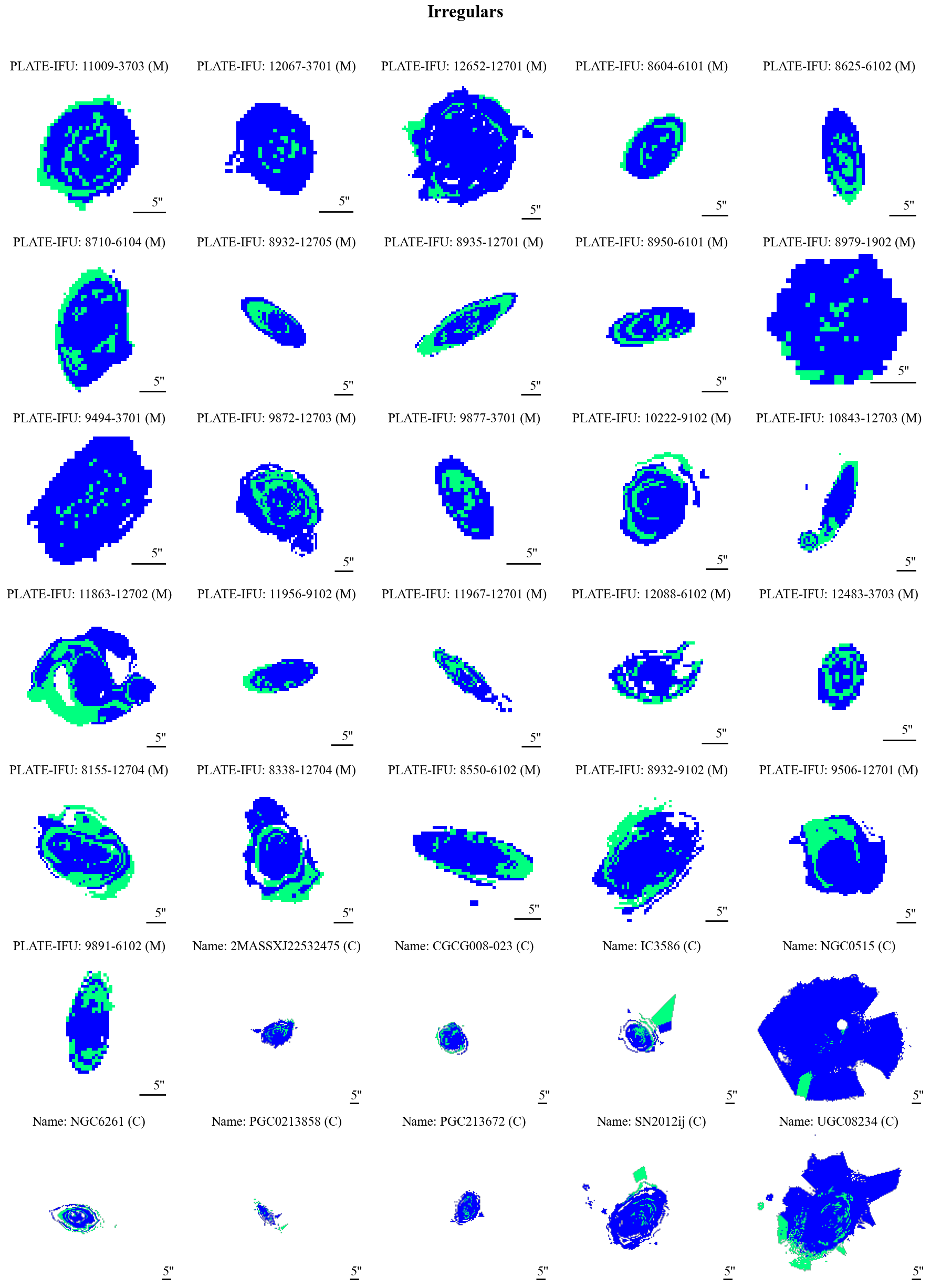}
\end{figure}
\clearpage
\begin{figure}
\centering
\includegraphics[width=\textwidth]{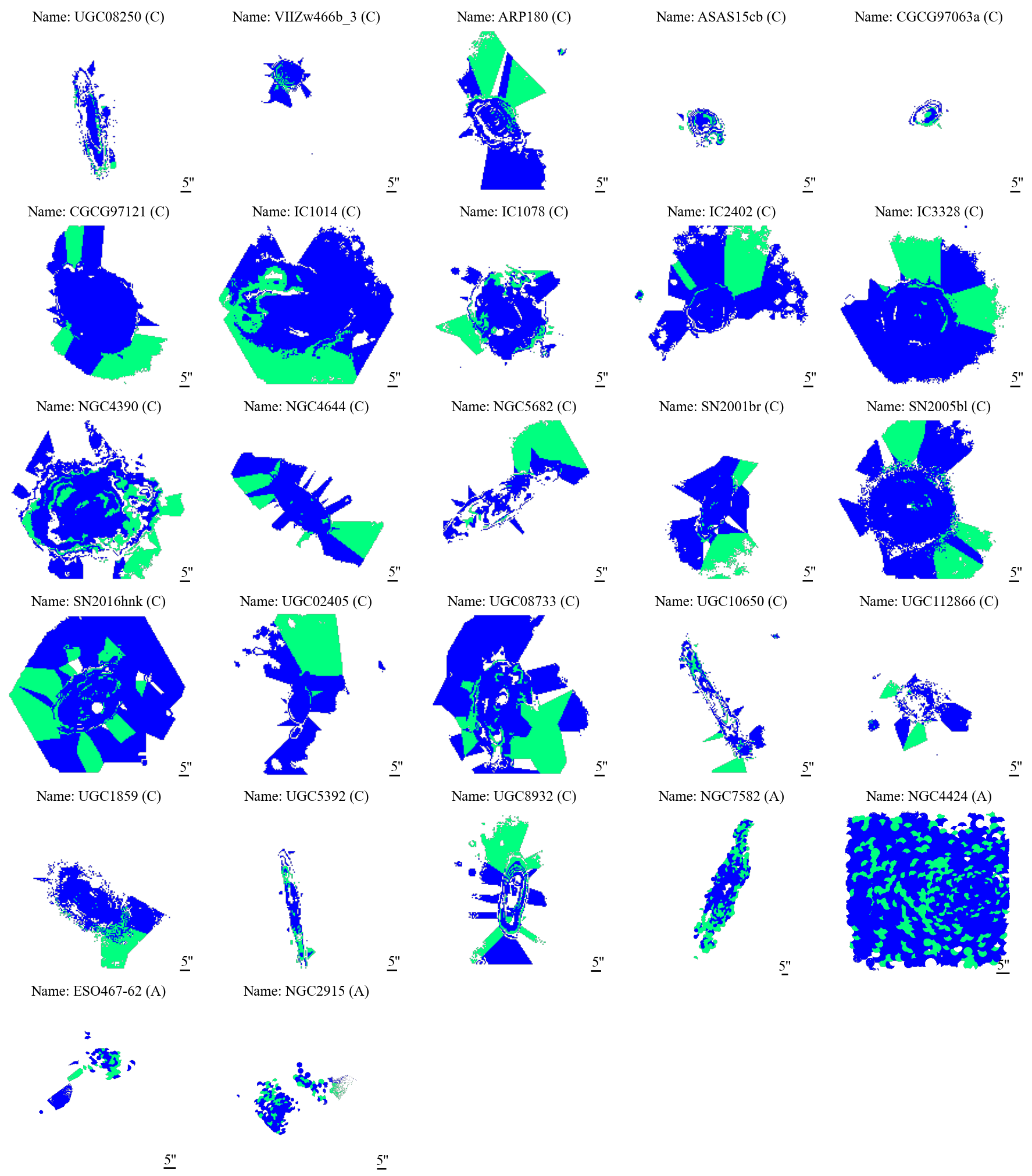}
\end{figure}
\clearpage
\begin{figure}
\centering
\includegraphics[width=\textwidth]{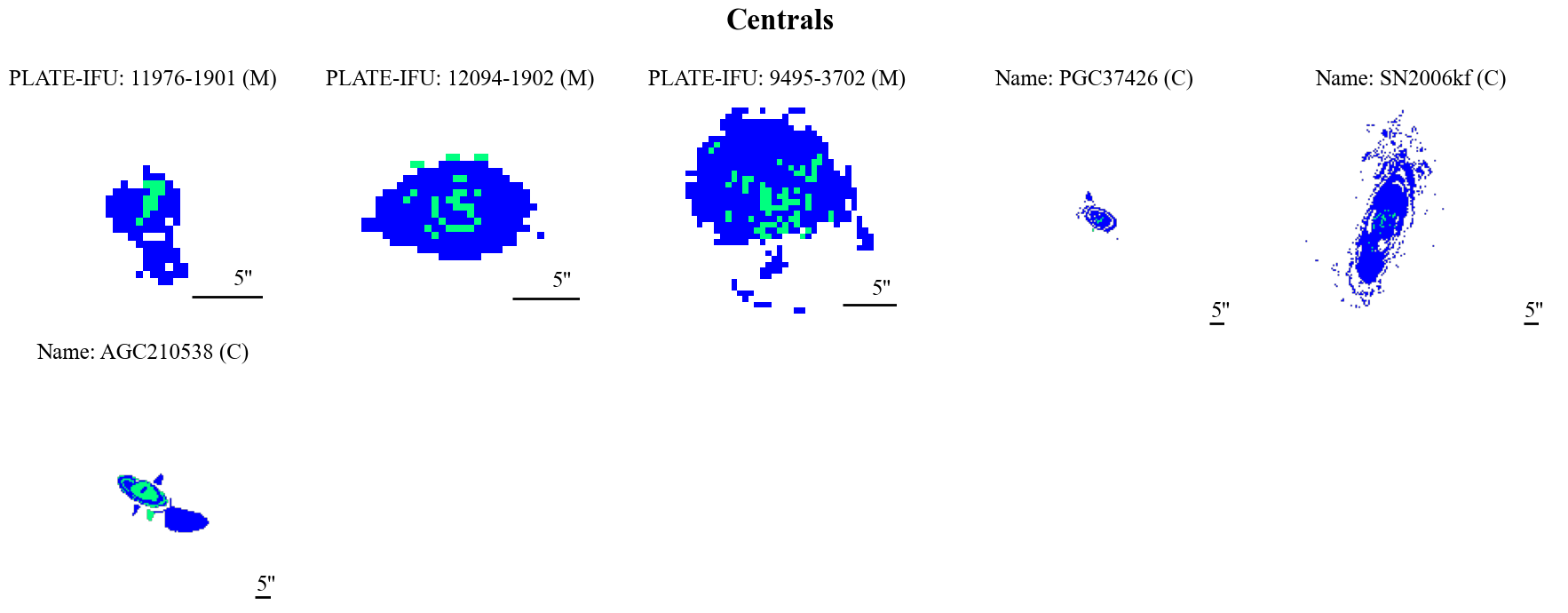}
\end{figure}
\clearpage
\begin{figure}
\centering
\includegraphics[width=\textwidth]{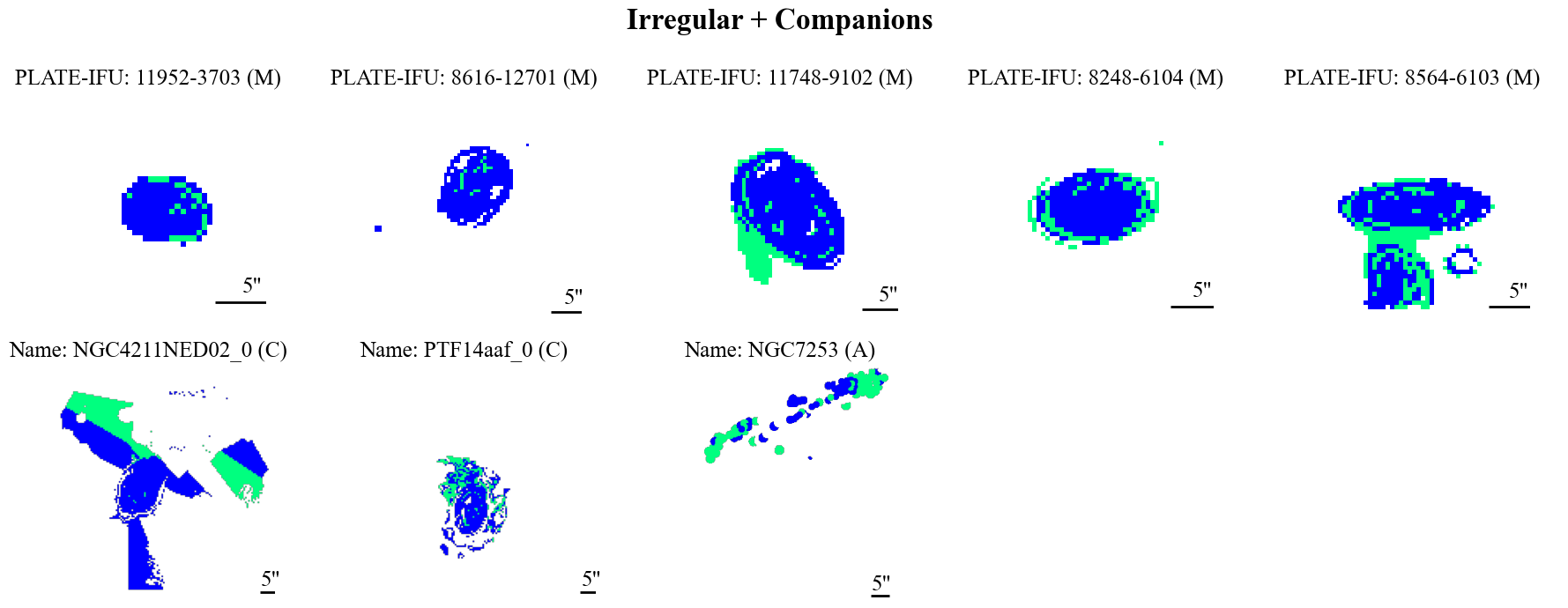}
\includegraphics[width=\textwidth]{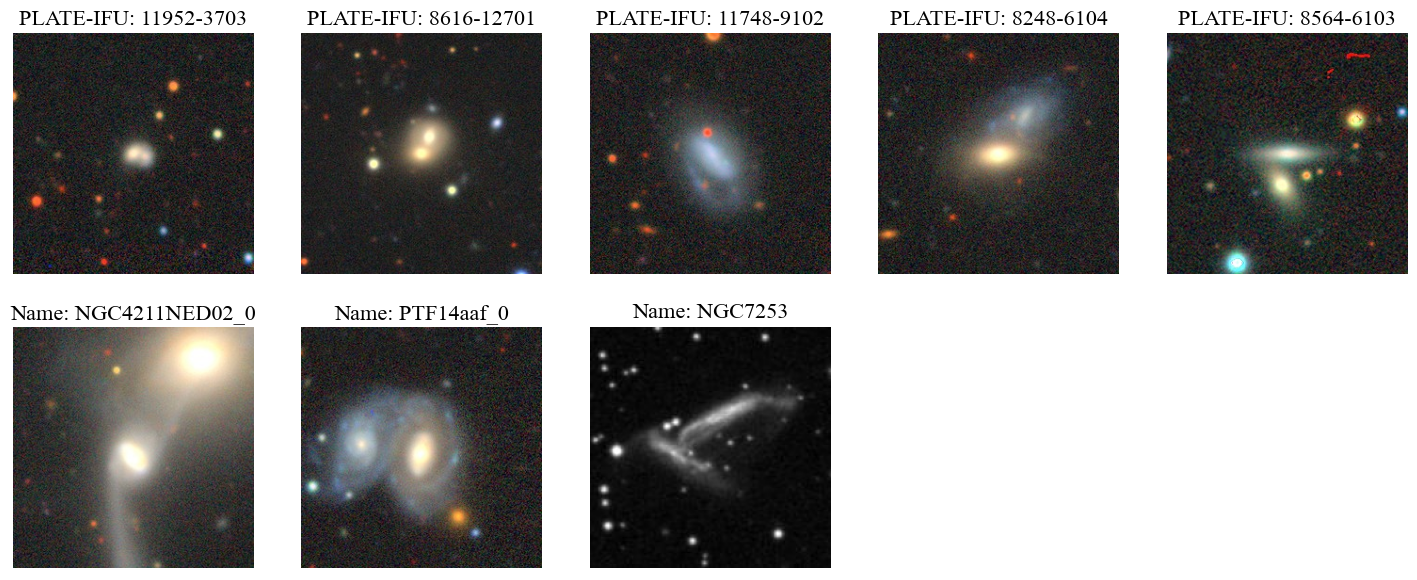}
\caption{Included here are the maps of the PSB spaxel distributions for all of our PSB galaxies. They are sorted into Rings, Irregulars, Centrals, and Irregular + Companions. According to our selection criteria, PSB spaxels and non-PSB spaxels are shown in green and blue, respectively. These maps are shown after the signal to noise cut. In the section of Irregular + Companion galaxies, the Legacy Survey images are shown below, since these were essential for determining the presence of merger features or a second galaxy (the DSS image is used in the case of the AMUSING++ galaxy). Parenthesis containing an M, C, or A are given after the galaxy identifiers, designating whether the parent survey was MaNGA, CALIFA, or AMUSING++, respectively.}
\label{fig:maps}
\end{figure}

\bibliography{sample631}{}
\bibliographystyle{aasjournal}

\end{document}